\documentclass[manuscript]{AGUJournal}
\journalname{JGR}

\usepackage{bm}
\usepackage{url}

\usepackage{graphicx}
\usepackage{amsmath}
\usepackage{amssymb}
\DeclareMathOperator{\Tr}{Tr}
\DeclareMathOperator{\diag}{diag}

\newenvironment{rev}{}{}

\begin{document}

\title{%
  On the Conditions of Aliasing Effects
  in Constellations
  with More than Four Spacecraft%
}

\authors{%
Lei Zhang%
\affil{1, 2, 3},
Jiansen He%
\affil{3},
Yasuhito Narita%
\affil{4, 5},
and Xueshang Feng%
\affil{2}%
}

\affiliation{1}{Qian Xuesen Laboratory Of Space Technology, Beijing, China, 100094}
\affiliation{2}{SIGMA Space Weather Group, National Space Sciences Center,
Chinese Academy Of Sciences, Beijing, China, 100190}
\affiliation{3}{School Of Earth And Space Sciences, Peking University, Beijing,
China, 100871}
\affiliation{4}{Space Research Institute, Austrian Academy of Sciences, Graz, Austria}
\affiliation{5}{%
Institut f\"ur Geophysik und Extraterrestrische Physik,
Technische Universit\"at Braunschweig, Braunschweig, Germany%
}

\correspondingauthor{Jiansen He}{jshept@pku.edu.cn}

\date{\today}

\begin{abstract}
The k-filtering/wave telescope method has been successfully applied
to the four-spacecraft measurements from Cluster II and MMS,
revealing the four-dimensional power spectral density $\textrm{PSD}(\omega, \mathbf k)$.
As the number of spacecraft within a constellation increases,
the geometry becomes more complicated,
and it is harder to find the first-Brillouin-like zone,
since we cannot use known formulas
to calculate the geometry of the zone.
To our knowledge, a quantitative study
on the aliasing effects as well as the
geometry of the first-Brillouin-like zone
for the constellation with more than 4 spacecraft
has not been proposed yet.
Herewith we provide a method to estimate and compare the geometry of
the first-Brillouin-like zone
in multi-spacecraft constellations.
The corresponding array of reciprocal lattice in the wavevector space
forms the Brillouin zone.
We set the preset PSDs
conforming to the ``slab-only'' and ``slab + 2D'' patterns
as the benchmark,
and compare them with the reconstructed PSDs
using 4, 5, and 9 virtual spacecraft.
\begin{rev}
Our results validates the estimation of the Brillouin-like zones.
\end{rev}
The method may provide a constraint of aliasing effects.
Our proposed method may be useful for the design and analysis
of future exploration sessions with multi-spacecraft constellation.
\end{abstract}

\section{Introduction}

The solar wind is an interplanetary plasma medium
abundant with turbulence\citep{tu95, brandenburg13},
and generally regarded as ``a natural and ideal turbulence laboratory''\citep{bruno13}.
Spectral index and spectral anisotropy provide
key hints for understadning the underlying processes of cascading and dissipation of turbulence
\citep{iroshnikov64, shebalin83, goldreich95, cho00, cho03, horbury08}.
One perception of turbulence,
both in interpretations of observed data and in simulation of turbulence,
regards the turbulence as a non-linearly-coupled wave packet
composed of various wave modes,
with components of magnetohydrodynamical (MHD) waves
(including Alfv\'en modes\citep{burlaga68, burlaga71, matthaeus96, wang12},
fast and slow modes\citep{cho02, cho03, vestuto03, yao11});
as well as kinetic waves \citep{sahraoui10prl, he11, klein12, salem12, narita14},
especially those with features of fluidization \citep{verscharen17}.

If the spectra about the wave vector $\mathbf{k}$ and the frequency $\omega$ can be estimated,
comparisons between theoretical and estimated dispersion relations can
separate wave modes and convective structures, or identify
wave modes.
(See~\citet{huang10, narita11, he12, shi15, wang16} etc.~for some previous endeavours.)
Observations testify spectral anisotropy
\citep{bieber96, maron01, horbury08, podesta09, chen10, forman11, turner12, he13, yan16}.
Owing to the complexity of solar wind and
the limitation of in-situ measurements,
there exists an intensive debate on the distribution of turbulence PSD in 3D wave-vector space.
Various types of 3D PSD distribution have been suggested theoretically and observationally
\citep{goldreich95, boldyrev06, mallet15, chen16}.
Some evidence is in favour of the ``slab + 2D'' scenario,
where the principal modes have quasi-parallel or quasi-perpendicular $\mathbf k$ \citep{horbury12, bruno13}.
For example, the distribution of spatial correlation function
in the 2D $(r_\parallel, r_\perp)$ space
for solar wind and magnetosheath turbulence illustrate
the pattern of ``slab + 2D'' at MHD scales and ion scales,
respectively \citep{matthaeus90, he11}.
The slab and 2D components at ion scales are found
to be consistent with the quasi-parallel ion cyclotron
waves and quasi-perpendicular Alfv\'en waves,
which together produce the magnetic helicity spectra
encompassing the two components \citep{he12, klein14}.

Multi-spacecraft observations provide important tools to measure
structures, waves, and turbulence in space plasma.
Missions Cluster II \citep{escoubet97, escoubet01} and MMS \citep{burch16, tooley2016} successfully provides data
sampled by a 4-spacecraft constellation,
which enables the separation of spatial and temporal changes,
as well as the development of 3D analysis methods
\citep{paschmann98, paschmann08}.
From observed physical quantities like magnetic field, velocity, or
number density of particles, these methods can be utilised to
estimate spatial derivatives, e.g.~current density\citep{dunlop90, chanteur93};
to find normal directions of structures, e.g.~minimum variable analysis\citep{sonnerup67, yao11, wang12};
and to detect $\mathbf{k}$ as a few peaks in spectral diagrams,
i.e.~filtering methods and singular value decomposition methods applied
to spectral matrices \citep{pincon91, santolik03}.

\begin{rev}
Wave telescope / k-filtering method\citep{neubauer90, motschmann96, glassmeier01}
mainly deduces phases from fluctuations of quantities like $n$ or $\mathbf{B}$.
It can quantitatively estimate power and provide
``a good presentation of the field turbulence'' \citep{pincon91}.
\end{rev}
Furthermore, physical relations between different fluctuating quantities,
e.g., the induction equation (Faraday's law),
are incorporated into the k-filtering method.
The geometric configuration of constellation greatly affects
the quality of wave-telescope analysis results.
For three-dimensional measurements,
the constellations themselves should be three-dimensional,
i.e.~not coplanar or collinear.
For 4-spacecraft constellations,
``tetrahedron geometric factors'' distinguish regular
(pyramids-like) tetrahedrons from irregular
(nearly coplanar or collinear) ones \citep{dunlop90, balogh97}.
For 5-spacecraft constellations,~\citet{pincon91} compared two cases
with differently located additional (the fifth) virtual spacecraft.
Adopting the wave telescope method to real data,
\citet{glassmeier01} introduced a ``Nyquist wave number''
to give an estimation of the wave vector range that
a constellation could resolve.
Empirically, \citet{sahraoui10} confined the scales of resolved waves as $[2d,10d]$,
where $d$ is the typical distance of spacecraft,
and the scales are affectd by spatial aliasing.
\begin{rev}
\citet{sahraoui10} also suggested to constrain the aliasing effects
by restricting the frequencies, i.e. to apply temporal filters.
\end{rev}

\begin{figure}[thb]
  \begin{center}
    \includegraphics[width=\textwidth]{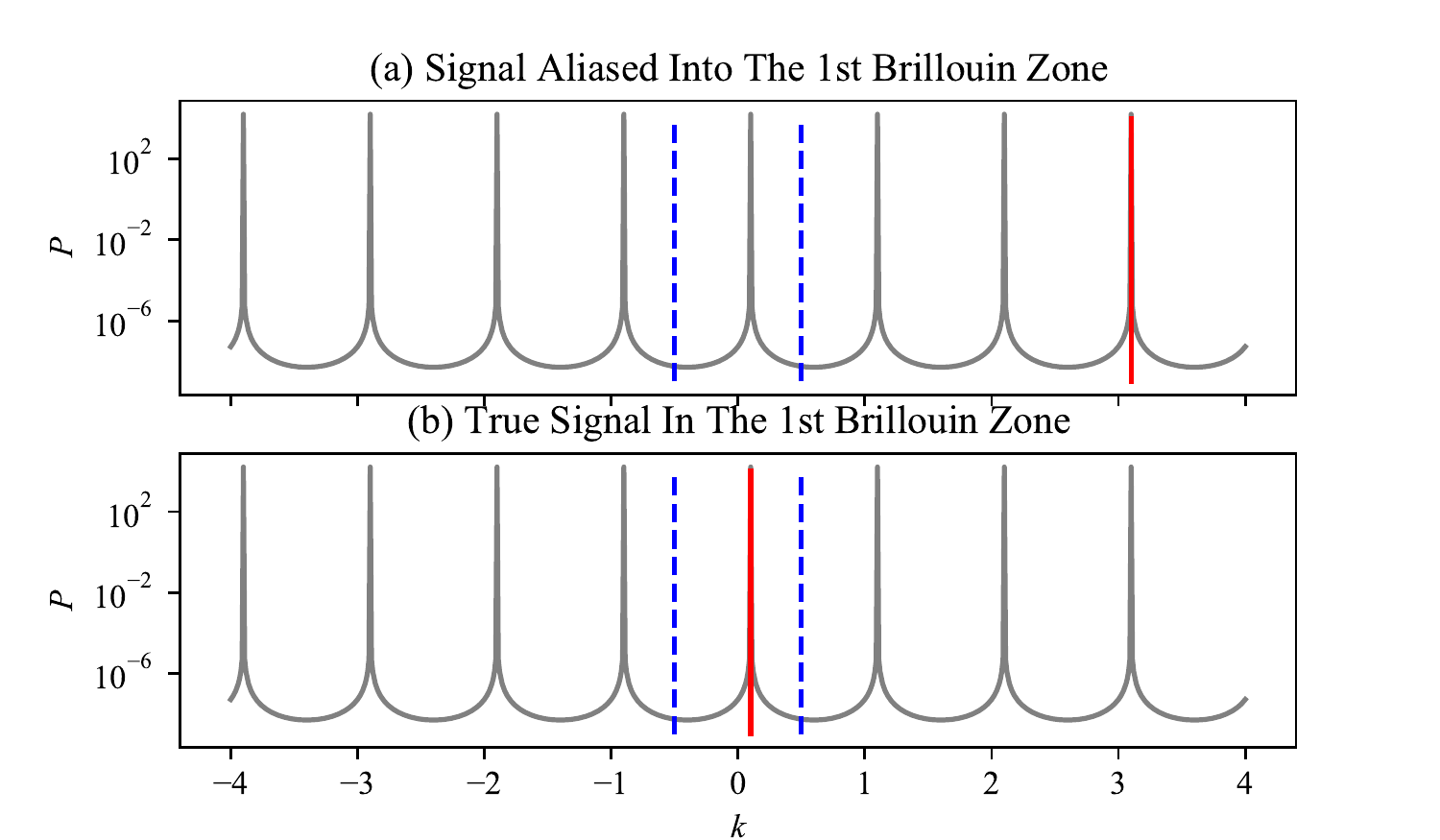}
  \end{center}
  \caption{%
  A simple 1D test about the concept of the first Brioullin zone:
  (a) the PSD profile in the zone is polluted by aliased signals from outside the zone;
  (b) the signals within the zone do not get aliased with each other.
  The reconstructed PSDs $P(k)$ are plotted in grey lines.
  The $k$ of the preset signals are marked with the red line.
  The first Brillouin zones are marked with the vertical blue dash lines.
  }\label{fig:simple_test}
\end{figure}

\begin{rev}
The aliasing effect is a long-lasting issue in spectral analysis
of space plasmas since the pioneering explorations
in the 1990s \citep{pincon88} and the early era of Cluster mission \citep{sahraoui03}.
\end{rev}
The spatial aliasing effects are usually inevitable,
because periodic discrete measurements of fluctuating physical quantities lead to corresponding
periodic pattern in Fourier domain
\citep[e.g.][]{neubauer90, tjulin05}.
The occurrence of aliasing effect can
distort the observed spectrum away from the real one\citep{narita09}.
Hence, it is important to calculate quantitatively the condition of aliasing effects.
If 2D or 3D constellations with even location of sampling points are available,
the discussion of aliasing effects can be conducted with the Nyquist-Shannon sampling theorem,
in analogy to its application to the cases of 1D uniform stencils.
However, the complex constellation with non-uniformally scattered spacecraft
in the 2D or 3D space hinders us from finding a simple definition of maximum wavenumber,
and the conditions of aliasing effects should be formulated directly from the $2n\pi$ difference in the phase.
\citet{neubauer90} and \citet{tjulin05} formulated the condition of aliasing effect, where wave-vectors $\mathbf k$ and $\mathbf k + \Delta \mathbf k$ are indistinguishable when
\begin{linenomath*}
\begin{equation}
  \Delta \mathbf k \cdot \mathbf x_{ij} = 2\pi n_{ij},
  \quad\textrm{for all } i, j \textrm{ such that }
  1 \le i < j \le N_{\textrm{s}},
  \label{eq:aliasing}
\end{equation}
\end{linenomath*}
where $N_{\textrm{s}}$ denotes the total number of spacecraft in the constellation,
and $n_{ij}$ are integers (can be zero).
  As \citet{neubauer90} stated, when there are only four spacecraft,
  the occurrences of aliasing effects in the $\mathbf k$-space form a lattice with parallelepiped cells defined by
  \begin{linenomath*}
  \begin{subequations}
    \begin{eqnarray}
      \Delta \mathbf k_1 &=& 2\pi(\mathbf x_{13} \times \mathbf x_{12}) / (\mathbf x_{14}\cdot(\mathbf x_{13}\times\mathbf x_{12})), \\
      \Delta \mathbf k_2 &=& 2\pi(\mathbf x_{14} \times \mathbf x_{12}) / (\mathbf x_{14}\cdot(\mathbf x_{13}\times\mathbf x_{12})), \\
      \Delta \mathbf k_3 &=& 2\pi(\mathbf x_{14} \times \mathbf x_{13}) / (\mathbf x_{14}\cdot(\mathbf x_{13}\times\mathbf x_{12})), \\
      \Delta \mathbf k &=&
        n_1\Delta\mathbf k_1 + n_2\Delta\mathbf k_2 + n_3\Delta\mathbf k_3,
    \end{eqnarray}
    \label{eq:solution4}
  \end{subequations}
  \end{linenomath*}
  where $n_1$, $n_2$, and $n_3$ are arbitrary integers.
  Using $\Delta\mathbf k_1$ to $\Delta\mathbf k_3$, one also defines a first Brillouin zone.
  \begin{rev}
  We should be aware that the concept and the condition of
  spatial aliasing effects are tricky to grasp.
  Even in the first Brillouin zone, the PSD profile can also be
  polluted by aliased signals with $\mathbf k$ outside the zone
  (see Figure \ref{fig:simple_test}a for a simple test).
  However, the concept of the Brillouin zones is stell essential:
  if there are two signals with both the wave vectors
  $\mathbf k_1$ and $\mathbf k_2$ located in the same Brillouin zone,
  then $\mathbf k_1$ does not get aliased with $\mathbf k_2$
  (see Figure \ref{fig:simple_test}b).
  \end{rev}

Eq.~\eqref{eq:solution4} perfectly describes the conditions of aliasing effects
  in a four-spacecraft constellation
theoretically.
  In a constellation with more than four spacecraft,
  from Eq.~\eqref{eq:aliasing} one cannot calculate
  $\Delta \mathbf k$ in a simple manner similar to
  Eq.~\eqref{eq:solution4}, because the equation becomes over-determined.
  For example, if there are 5 spacecraft, Eq.~\eqref{eq:solution4} will consist of 4 independent equations,
  but only 3 components are to be determined.
Therefore, a method to estimate
the Brillouin-like zone should be developed for constellations with more than 4 spacecraft.

Herewith we aim our research at the aliasing effect,
and propose a practicalness-oriented method to quantitatively estimate the wave vector limitations.
This shows our attempts of exploration about how different configurations
of constellations can help constrain the aliasing effects better.
To illustrate the application of this method, we also benchmark-test
the constellations with different numbers of spacecraft.
Section 1 introduces the topics of turbulence, multi-spacecraft observations
and methodology, PSD reconstructions, and the aliasing effect in detail.
Section 2 briefly formulates the MSR method and settings of the synthetic data.
Section 3 demonstrates our proposed practical method,
which helps to successfully identify the first-Brillouin-like zone.
Section 4 shows our benchmark tests with virtual constellations.
Section 5 summarizes the research and discusses some issues of our methods.

\section{Formulation of Filtering Methods and Their Tests}

For an example of filtering methods, we choose the Multi-point Signal Resonator (MSR) method
\citep{samson83, samson83b, schmidt86, choi93, narita11}
for the benchmark test to reconstruct the 4D PSD of synthetic turbulence-like data.
Some basic settings of test signal generators are also briefly introduced.
Finally, the geometric configurations of virtual constellation are also presented here.

\subsection{MSR Method}\label{subsect:msr}

The MSR method operates on several simultaneous observations,
and it aims at resolving peaks of wave signals in power spectra
when a moderate number of waves co-exist in a frequency band.
The actual filtering has two parts: Capon filter \citep{capon69, capon69geoph, capon69jgr} and
MUltiple SIgnal Classification (MUSIC) \citep{samson83, samson83b, narita11}.
The former is intended for finding the wave signals and estimating their power;
the latter is employed to reduce further the noise.
As expected, the ``total'' MSR filter is composed as the product of the Capon and the MUSIC parts,
i.e.~multiplying them together,
and thus it gives the final estimation of PSD\@.

Here begins the formulation of the MSR method.
Samplings of the data occur when $t_j=j\Delta{t}$, $j=1,2,\ldots,N_{\mathrm{t}}$,
with a uniform cadence $\Delta{t}$ till $N_{\mathrm{t}}$ samples.
We also assume that all the $N_{\mathrm{s}}$ spacecraft do not change their positions.
Hence the $i$-th spacecraft's $j$-th sample of the field $\mathbf{B}(\mathbf{x},t)$ is
$\mathbf{B}_{ij}=\mathbf{B}(\mathbf{x}_i,t_j)$.

The MSR technique minimizes noise in the $\mathbf{k}$-space.
The filtering is calculated using a set of test wave-vectors $\mathbf{k}'$.
At first, spatial effects must be separated from temporal effects.
Therefore, a Fourier transform is applied along $t$,
where the Fourier-transformed image $\tilde{\mathbf{B}}_{ij}$
has a frequency $\omega_j=2\pi{j}/(N_{\textrm{t}}\Delta{t})$.

With the aforementioned $\tilde{\mathbf B}_{ij}$,
the two parts starts with the vector $\mathbf{S}_j$ , which is constructed as
\begin{linenomath*}
\begin{equation}
  \mathbf{S}_j={({B_x}_{1j},{B_y}_{1j},{B_z}_{1j},{B_x}_{2j},\ldots,{B_z}_{N_{\textrm{s}}j})}^{\mathrm{T}}.
\end{equation}
\end{linenomath*}
The Capon part estimates the power at a `test wave-vector' $\mathbf{k}'$
{{(i.e.~the pre-assumed wave-vector to build filters)}}
as
\begin{linenomath*}
\begin{equation}
  P_{\textrm{Capon}}(\mathbf{k}')=\frac{1}{\Tr{(\mathsf{V}^{\dagger}\cdot\mathsf{H}^{\dagger}\cdot\mathsf{R}^{-1}\cdot\mathsf{H}\cdot\mathsf{V})}}.
  \label{eq:Capon}
\end{equation}
\end{linenomath*}
where $\mathsf{R}=\mathbf{S}^{\dagger}\mathbf{S}$,
also with $\mathsf{H}$ and $\mathsf{V}$ defined as
\begin{linenomath*}
\begin{subequations}
  \begin{eqnarray}
    \mathsf{H}(\mathbf{k}')&=&\begin{pmatrix}\mathsf{P}_1& \mathsf{P}_2&\cdots&\mathsf{P}_{N_\textrm{s}}\\\end{pmatrix}^{\mathrm{T}},\\\mathsf{P}_i&=&(\exp{(-{\mathrm{i}}\mathbf{k}'\cdot{\mathbf{x}}_i)})\cdot\mathsf{I}=\begin{pmatrix}{\mathrm{e}}^{-{\mathrm{i}}\mathbf{k}'\cdot{\mathbf{x}}_i}&0&0\\0&{\mathrm{e}}^{-{\mathrm{i}}\mathbf{k}'\cdot{\mathbf{x}}_i}&0\\0&0&{\mathrm{e}}^{-{\mathrm{i}}\mathbf{k}'\cdot{\mathbf{x}}_i}\\\end{pmatrix},\\\mathsf{V}(\mathbf{k}')&=&\mathsf{I}+\hat{\mathbf{k}'}^{\textrm{T}}\hat{\mathbf{k}'}=\begin{pmatrix}1+\frac{{k'_x}^2}{{k'}^2}&\frac{k'_xk'_y}{{k'}^2}&\frac{k'_xk'_z}{{k'}^2}\\\frac{k'_yk'_x}{{k'}^2}&1+\frac{{k'_y}^2}{{k'}^2}&\frac{k'_yk'_z}{{k'}^2}\\\frac{k'_zk'_x}{{k'}^2}&\frac{k'_zk'_y}{{k'}^2}&1+\frac{{k'_z}^2}{{k'}^2}\\\end{pmatrix}.
  \end{eqnarray}
\end{subequations}
\end{linenomath*}
Here \citet{motschmann96} introduced the matrix $\mathsf V$
to enforce the restriction $\nabla \cdot \mathbf B = 0$.

Theoretically, since it conforms to~\citet{pincon91}'s
general form of filtering methods,
the Capon method can spot the wave signals and calculate their power.
However, its resolution at peaks is rather limited \citep{narita11},
and the MUSIC part is designed to suppress noise levels and improve resolutions
by multiplying a ``filter''
\begin{linenomath*}
\begin{equation}
  P_{\textrm{M}}(\mathbf{k}')=\frac{1}{{P_{\textrm{M}}}_{\max{}}}\frac{1}{\Tr{(\mathsf{V}^{\dagger}\cdot\mathsf{H}^{\dagger}\cdot\mathsf{F}\cdot\mathsf{\Lambda}\cdot\mathsf{F}^{\dagger}\cdot\mathsf{H}\cdot\mathsf{V})}}
\end{equation}
\end{linenomath*}
to $P_{\textrm{Capon}}$ defined by Eq.~\eqref{eq:Capon}.
Denote the eigenvalues of $\mathsf{R}$ as $\lambda_1,\lambda_2,\ldots,\lambda_N$ (here $N=3N_\textrm{s}$) where $\lambda_1\le\lambda_2\le\ldots\le\lambda_N$,
and the corresponding eigen-vectors as $\mathbf{e}_1,\mathbf{e}_2,\ldots,\mathbf{e}_N$.
The matrix $\mathsf{F}$ is constructed as
\begin{linenomath*}
\begin{equation}
  \mathsf{F}=\begin{pmatrix}\mathbf{e}_1&\mathbf{e}_2&\cdots&\mathbf{e}_N\\\end{pmatrix},
\end{equation}
\end{linenomath*}
and the matrix $\mathsf{\Lambda}$ as
\begin{linenomath*}
\begin{equation}
  \mathsf{\Lambda} = \diag\left({\left(\dfrac{\lambda_1}{\lambda_N}\right)}^{-n},{\left(\dfrac{\lambda_2}{\lambda_N}\right)}^{-n},\ldots{\left(\dfrac{\lambda_N}{\lambda_N}\right)}^{-n}\right)
\end{equation}
\end{linenomath*}
with $n=2$.
The maximum ${{P_{\textrm{M}}}_{\max{}}}$ is defined as
\begin{linenomath*}
\begin{equation}
  {{P_{\textrm{M}}}_{\max{}}}=\max_{\mathbf{k}', \omega}P_{\textrm{M}}(\mathbf{k}', \omega).
\end{equation}
\end{linenomath*}

Some cautions must be exercised to apply this method numerically.
\begin{rev}
The matrix $\mathsf{R}$ is singular,
so we must replace $\mathsf{R}$ with
$(1-\epsilon)\mathsf{R}+\epsilon\mathsf{I}$.
\end{rev}
Here we take $\epsilon=10^{-8}$.
Meanwhile, in order to provide enough degrees of freedom and
to make $\mathsf{R}$ more statistically significant (e.g.~\citet{narita11}),
in actual filters,
$\mathsf{R}$ must be averages of FFTs of many (24 here) segments of $\mathbf{B}_{ij}$.
  The results are not significantly sensitive to the number of segments,
  if the number is big enough.

\subsection{Generation of Turbulence-like Signals }\label{subsect:wavepacket}
The formula for a packet of waves is expressed as
\begin{linenomath*}
\begin{equation}
  \mathbf{B}(\mathbf{x},t)=\mathbf{B}_0 + \sum_{j}\mathbf{B}_j\cos(\mathbf{k}_j\cdot\mathbf{x}-\omega_j{t}+\phi_j),\label{wavepacket}
\end{equation}
\end{linenomath*}
where $j$ denotes a wave with a real-valued amplitude ${\mathbf{B}}_j$,
an initial phase $\phi_j$, a wave-vector $\mathbf{k}_j$,
and an angular-frequency $\omega_j$.
The waves are superposed on a background magnetic field $\mathbf B_0$.

To simulate turbulence in the solar wind,
we consider the following issues in designing the waves.
Since Alfv\'enic mode dominates,
oscillations obey the dispersion and polarization relations
of Alfv\'en waves:
\begin{linenomath*}
\begin{subequations}
  \begin{eqnarray}
    \omega_j&=&v_{\mathrm{A}}\hat{\mathbf{B}_0}\cdot\mathbf{k}_j+\mathbf{v}_{\textrm{SW}}\cdot\mathbf{k}_j,\label{eqn:dispersion}\\{\mathbf{B}}_j&\parallel&(\mathbf{B}_0\times\mathbf{k}_j).
  \end{eqnarray}
\end{subequations}
\end{linenomath*}
Here $\mathbf{B}_0$ is a preset background field,
$v_{\mathrm{A}}$ is the Alfv\'en speed,
and $\mathbf{v}_{\textrm{SW}}$ is the velocity of solar wind flow.
  Here every quantity is defined and measured in the reference frame of spacecraft, in order to simplify discussion and to skip conversions.
  The term in Eq.~\eqref{eqn:dispersion} stems from the dispersion relation
  $\omega'_j = v_{\mathrm{A}} \hat{\mathbf{B}_0}\cdot\mathbf{k_j}$,
  just after considering the Doppler-shift effect.

In this investigation,
we set $\mathbf{B}_0$ as $(10/\sqrt3,10/\sqrt3,10/\sqrt3)$,
$\mathbf{v}_{\textrm{SW}}=(0,0,7\sqrt3/2)$,
and $v_{\mathrm{A}}=1$.
Thus $\mathbf{B}_0$ and $\mathbf{v}_{\textrm{SW}}$ forms an angle of
$54.7^\circ$, conforming to the typical range for the solar wind at 1AU\@.
When $\mathbf{k}\parallel\mathbf{B}_0$, the polarization of
magnetic field $\mathbf{B}_j$ is randomly chosen,
with considerations to keep $\mathbf{B}_j\perp\mathbf{k}$.
In this way, $\mathbf{k}_j$ decides most properties of a wave component,
so it must be carefully set.
The detailed information, as involved with the slab only / slab + 2D scenario,
will be fully given out in Sect.~4.
Finally, the initial phase $\phi_j$ is randomly chosen in $[0, 2\pi]$.

Such settings need some explanations.
Firstly, we use only one set of directions of $\mathbf{B}_0$ and $\mathbf{v}_\textrm{SW}$.
Though a variety of settings might cover various statuses of the solar wind,
one set is sufficient for a simplified benchmark.
Secondly, we choose to formulate our benchmark in a dimensionless manner to simplify the discussion.
For reference, the basis of normalization can be assumed with typical parameters in the solar wind at 1AU:
e.g.~for magnetic field strength, one unit equals to 1~nT;
for velocity, one unit as $50\;\textrm{km}\,\textrm{s}^{-1}$;
for $\mathbf{x}_i$, a typical spatial scale might be assumed (e.g. $10^4\;\textrm{km}$).
One unit of time is determined accordingly.

\subsection{Configuration of Virtual Spacecraft Constellations}

\begin{figure}[htbp]
  \begin{center}
    \includegraphics[width=\textwidth]{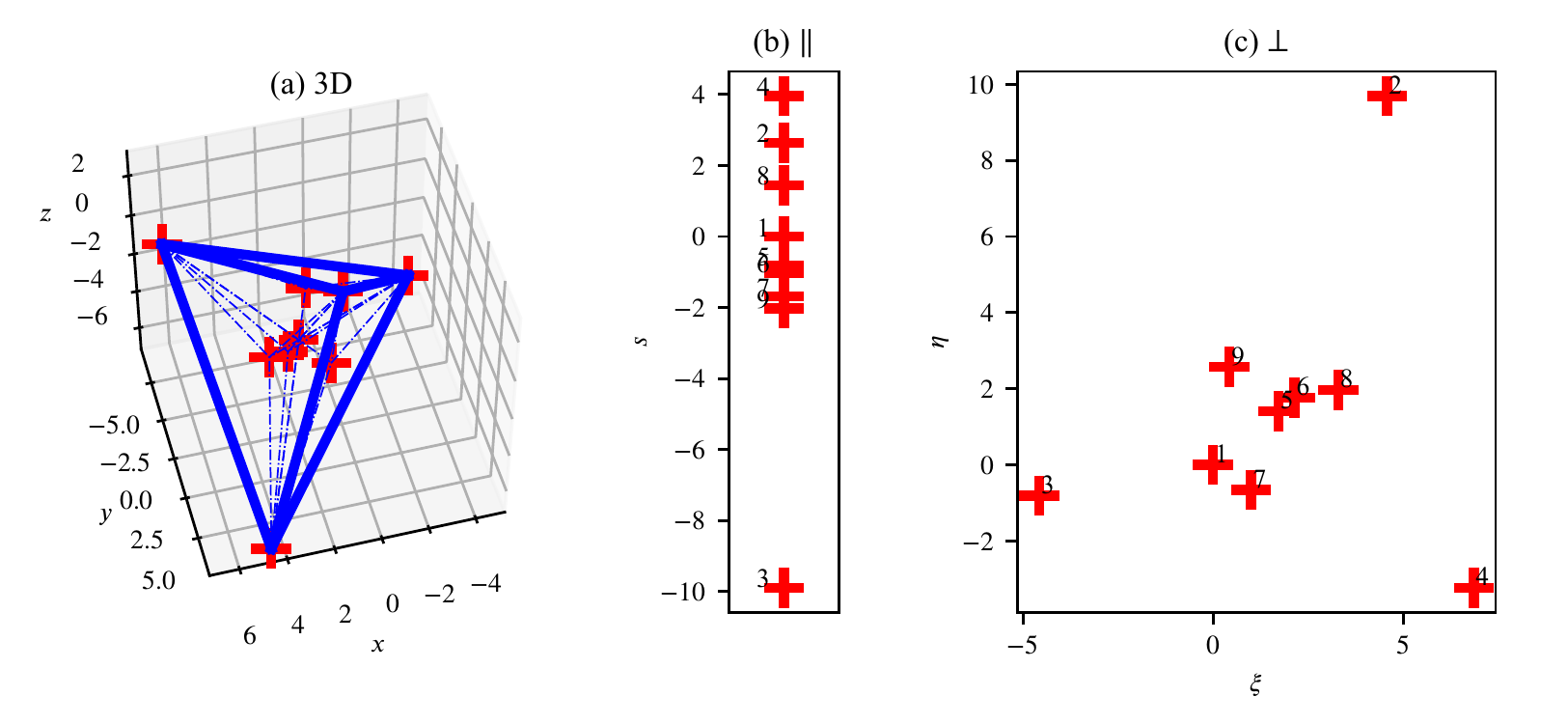}
  \end{center}
  \caption{%
  The locations of the virtual spacecraft.
  (a) 3D-vision. Each cross symbol in red denotes one spacecraft.
  The bold blue lines show the ``basic tetrahedron''.
  Thin blue lines show the additional tetrahedrons.
  (b) Projections along the $\mathbf{B}_0$ direction. Here the coordinate $s$ indicates the coordinate projected.
  (c) Projections onto the plane perpendicular to $\mathbf{B}_0$.
  Here the coordinates $\xi$ and $\eta$ denote two axes thereon.
  }\label{fig:spacecraft}
\end{figure}

In this benchmark test, to highlight improvements
brought by constellations with more spacecraft,
three constellations of 4, 5, and 9 spacecraft are formed.
The designed locations are
$\mathbf{x}_1,\mathbf{x}_2,\ldots,\mathbf{x}_9$.
The three constellations are composed of $\mathbf{x}_1$ to $\mathbf{x}_4$, $\mathbf{x}_1$
to $\mathbf{x}_5$, and $\mathbf{x}_1$ to $\mathbf{x}_9$, respectively.
In this test, we take
\begin{linenomath*}
\begin{subequations}
  \begin{eqnarray}
    \mathbf{x}_5&=&\frac{\mathbf{x}_1+\mathbf{x}_2+\mathbf{x}_3+\mathbf{x}_4}{4},\\\mathbf{x}_6&=&\frac{\mathbf{x}_5+\mathbf{x}_2+\mathbf{x}_3+\mathbf{x}_4}{4},\\\mathbf{x}_7&=&\frac{\mathbf{x}_1+\mathbf{x}_5+\mathbf{x}_3+\mathbf{x}_4}{4},\\\mathbf{x}_8&=&\frac{\mathbf{x}_1+\mathbf{x}_2+\mathbf{x}_5+\mathbf{x}_4}{4},\\\mathbf{x}_9&=&\frac{\mathbf{x}_1+\mathbf{x}_2+\mathbf{x}_3+\mathbf{x}_5}{4},
  \end{eqnarray}
\end{subequations}
\end{linenomath*}
i.e.~every additional spacecraft is located at the barycentre of a designed
tetrahedron. Therefore, the tetrahedron of $\mathbf{x}_1$ to $\mathbf{x}_4$
is so fundamental that we name it the ``basic tetrahedron''.
  In this benchmark test, in order to show aliasing effects,
  we design the Brillouin zone by the 4-spacecraft constellation,
  and use Eq.~\eqref{eq:aliasing} to solve
  the corresponding locations of the spacecraft.
  The solution is made by assigning $\Delta \mathbf{k}_1$ to
  $\Delta \mathbf{k}_3$ \textit{a priori},
  and compute
  \begin{linenomath*}
  \begin{subequations}
  \begin{eqnarray}
    \mathbf{a}_1 &=& \Delta \mathbf{k}_2 \times \Delta \mathbf{k}_3, \\
    \mathbf{a}_2 &=& \Delta \mathbf{k}_3 \times \Delta \mathbf{k}_1, \\
    \mathbf{a}_3 &=& \Delta \mathbf{k}_1 \times \Delta \mathbf{k}_2.
  \end{eqnarray}
  \end{subequations}
  \end{linenomath*}
  Then we assign
  \begin{linenomath*}
  \begin{subequations}
  \begin{eqnarray}
    \mathbf{x}_1 &=& (0, 0, 0), \\
    \mathbf{x}_2 &=& \frac{2\pi\mathbf{a}_1}{\mathbf{a}_1\cdot\Delta\mathbf{k}_1}, \\
    \mathbf{x}_3 &=& \frac{2\pi\mathbf{a}_2}{\mathbf{a}_2\cdot\Delta\mathbf{k}_2}, \\
    \mathbf{x}_4 &=& \frac{2\pi\mathbf{a}_3}{\mathbf{a}_3\cdot\Delta\mathbf{k}_3}.
  \end{eqnarray}
  \end{subequations}
  \end{linenomath*}

  Here we design
  \begin{linenomath*}
  \begin{subequations}
  \begin{eqnarray}
    \Delta\mathbf{k}_1 &=& (\frac14, \frac14, -\frac12), \\
    \Delta\mathbf{k}_2 &=& (0, -\frac12, -\frac12), \\
    \Delta\mathbf{k}_3 &=& (\frac34, -\frac12, 0).
  \end{eqnarray}
  \label{eq:designed_k}
  \end{subequations}
  \end{linenomath*}
  Hence we get a solution
  \begin{linenomath*}
  \begin{subequations}
  \begin{eqnarray}
    \mathbf x_1 &=& (0, 0, 0), \\
    \mathbf x_2 &=& \frac{4\pi}{11} \cdot (4, 6, -6), \\
    \mathbf x_3 &=& \frac{4\pi}{11} \cdot (-4, -6, -5), \\
    \mathbf x_4 &=& \frac{4\pi}{11} \cdot (6, -2, 2).
  \end{eqnarray}
  \end{subequations}
  \end{linenomath*}
The locations are plotted in Figure~\ref{fig:spacecraft}.
In this figure, in order to describe the slab + 2D scenario,
we introduce a new coordinate system $(s, \xi, \eta)$.
The $s$-axis aligns along ${\mathbf B}_0$.
The other two axes align in the perpendicular plane.
This coordinate system is not used in actual computation,
but provides some help in discussions on the geometric configuration of the constellations.

\section{Practical Calculation of Occurrences of Aliasing Effects}

  To illustate the necessity and the difficulties involved,
  we analyse a 1D problem in detail.
Then we present our practical methods in the 3D geometry,
  and apply our methods to an MMS constellation configuration.

\subsection{Identifying Aliasing Effects in 1D Cases}
\begin{figure}[htbp]
  \begin{center}
    \includegraphics[width=\textwidth]{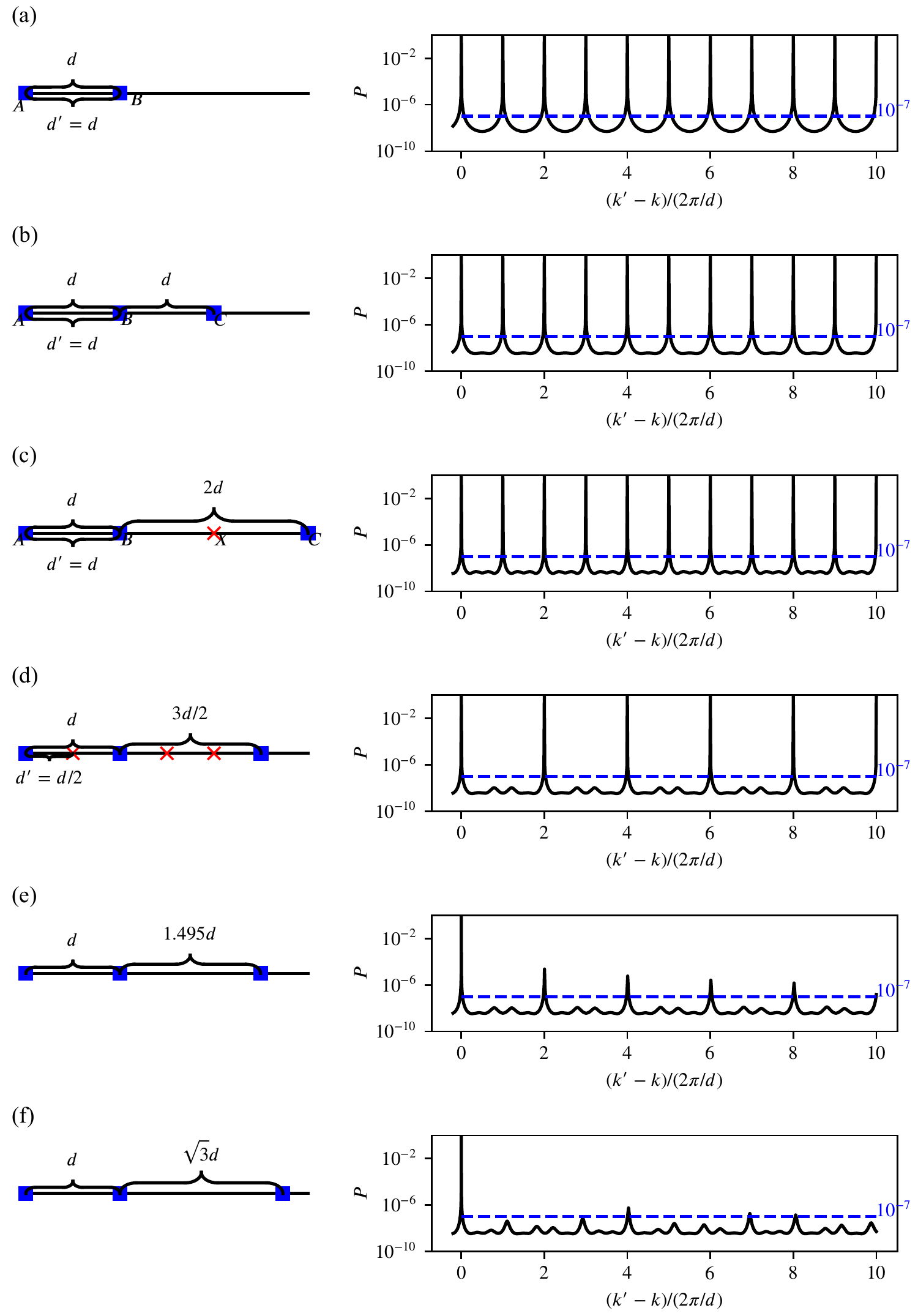}
  \end{center}
  \caption{%
    One-dimensional cases: locations of virtual spacecraft (left)
    and their aliasing effects (right).
    Spacecraft are marked with blue squares,
    and uniform stencil locations are marked with red crosses if no spacecraft is present.
    The typical scale $d$ and its corresponding distance $d'$ (if available) are also marked.
    The plots in the right column show the power $P$
    estimated by the Capon filter versus $\Delta k = k' - k$,
    normalised by $(2\pi/d)$.
    Panels (a) to (f) show each cases.
  }\label{fig:case-1d}
\end{figure}

First we consider a 1D problem before investigating complicated 3D cases.
In this problem, we set up a few configurations of virtual
spacecraft and check the aliasing effects occurring therein
(see Figure~\ref{fig:case-1d}).
For simplicity, here we use a Capon filter directly.

We start with Case (a), where two virtual spacecraft are located
with a distance $d$.
The aliasing effects occur when $k' - k = n \cdot (2\pi / d)$,
from Eq.~\eqref{eq:aliasing}.
Then we introduce the additional third spacecraft at different locations.
In Case (b), the spacecraft are aligned so that $AB = d$ and $BC = d$.
The aliasing-free range of $k$ does not increase.
In Case (c), the distance $BC$ increases to $2d$,
but the aliasing effects still occur in the same condition.
The absence of improvements can be explained
either directly with Eq.~\eqref{eq:aliasing} or with the Nyquist-Shannon theorem:
if we add an imaginary probe at the middle of $BC$ (denoted as $X$),
it is reasonable to assign a uniform distance $d' = d$.
\begin{rev}
Here $d'$ is defined as the maximum of all the distances $d''$,
where $AB$ and $BC$ are both multiples of $d''$.
\end{rev}
In the sense of $d'$, Case (c) does not differ from Cases (a) or (b).
To have a greater aliasing-free range of $k$,
we adjust $BC$ in Case (d) to $3d/2$.
Here we can take $d' = d/2$,
and both Eq.~\eqref{eq:aliasing} and the Nyquist-Shannon theorem
predict that the aliasing effects occur when $k' - k = n \cdot (4\pi / d)$.

However, in practical cases, the spacecraft may not form an array
where $d'$ is exactly defined.
In Case (e), $BC$ differs slightly from in Case (d).
If we treat 1.495 as the exact number $299/200$ and thus calculate $d' = d/200$,
a strict solution of Eq.~\eqref{eq:aliasing} gives $k' - k = 2\pi n / d' = n \cdot (400 \pi / d)$,
but the plot of $P(k')$ reveals the great pollution of the reconstructed spectrum at $k' - k = 4\pi / d$.
Therefore, in practical cases, aliasing effects should also be detected practically,
and the threshold might be set to a small value.
In Case (f), we even cannot form a uniform stencil and define an exact distance $d'$.
However, the aliasing effects occur;
if we take $10^{-7}$ as a threshold, it appears at $(k' - k) / (2\pi/d) \approx 3$,
4, 7, 8 etc.

\subsection{Identification of First-Brillouin-like Zones in 3D Geometry}

\begin{figure}[t]
  \begin{center}
    \includegraphics[width=\textwidth]{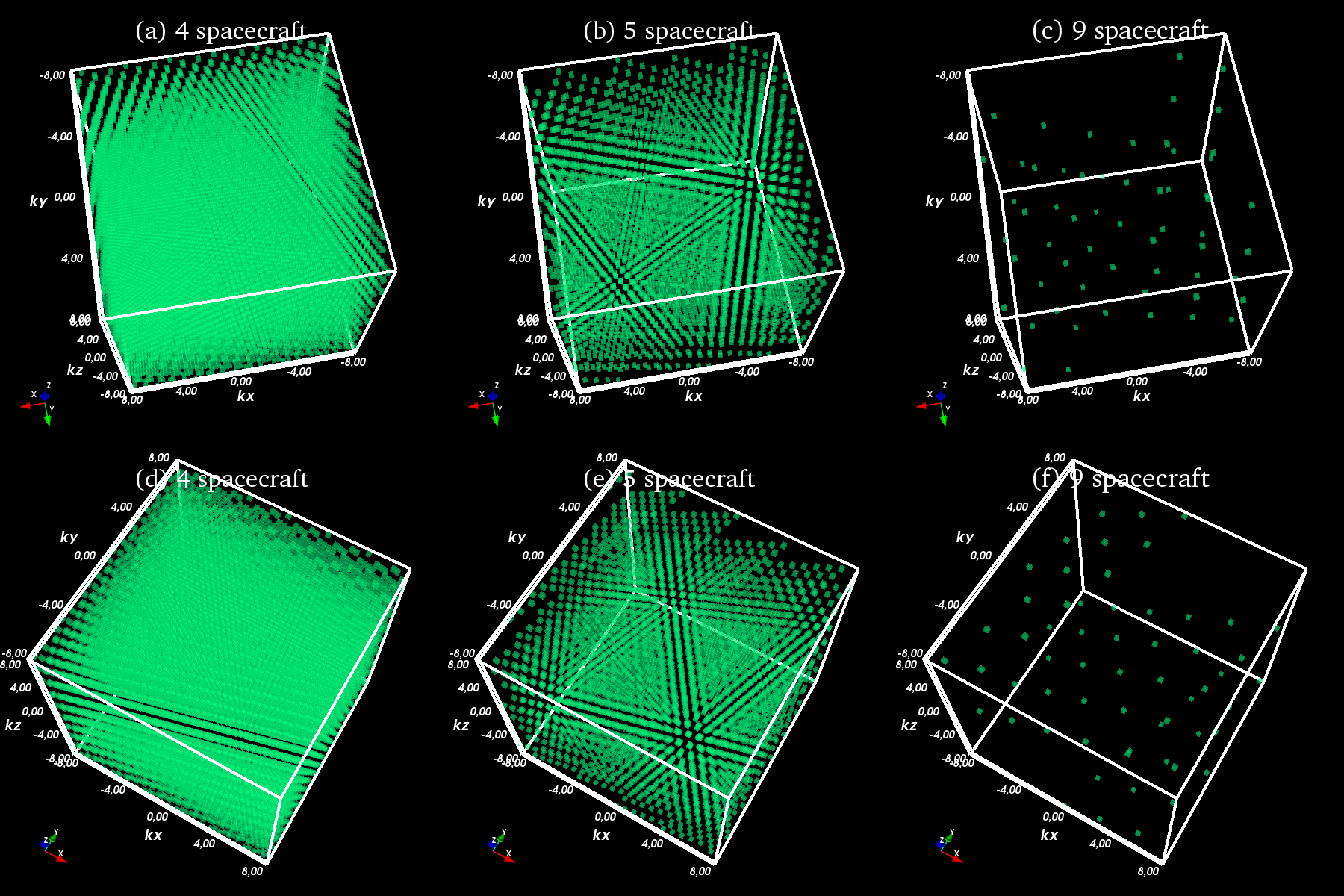}
  \end{center}
  \caption{%
  Aliasing effects as estimated using Capon filters only.
  Subplots (a) to (c) respectively correspond to
  the constellations with 4, 5, and 9 virtual spacecraft,
  whose locations are described in Sect.~2.3.
  Subplots (d) to (f) are essentially the same as (a) to (c),
  but plotted from another viewing point.
  }\label{fig:solved-aliasing}
\end{figure}

\begin{figure}[t]
  \begin{center}
    \includegraphics[width=\textwidth]{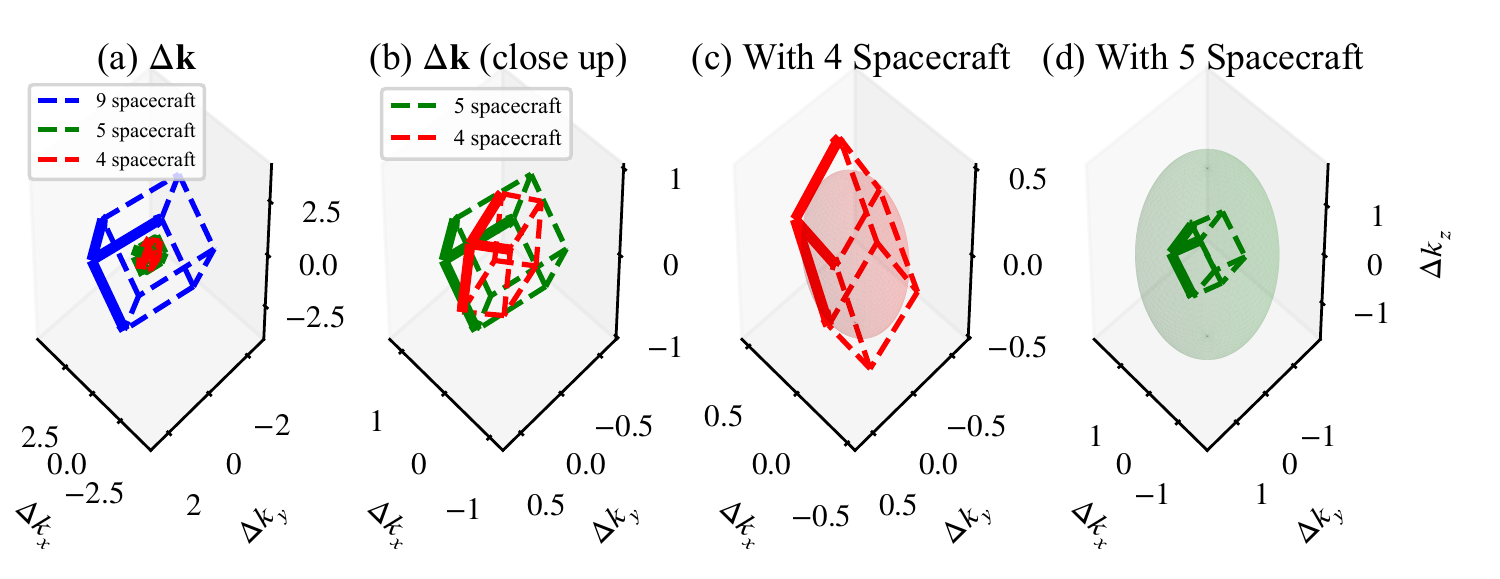}
  \end{center}
  \caption{%
    Comparisons of occurrences of aliasing effects.
    (a) $\Delta \mathbf k$ of the constellations with
    9(blue), 5(green), and 4(red) spacecraft.
    (b) A close up of (a) with only the 5(green) and 4(red)
    spacecraft cases.
    (c) The comparison of the minimal $\Delta \mathbf k$
    cell and the spherical zone $k \le \pi / d_{\min{}}$
    as in the 4-spacecraft constellation.
    (d) The same comparison as (b), but in the 5-spacecraft constellation.
  }\label{fig:compare-459}
\end{figure}

\begin{figure}[t]
  \begin{center}
    \includegraphics[width=\textwidth]{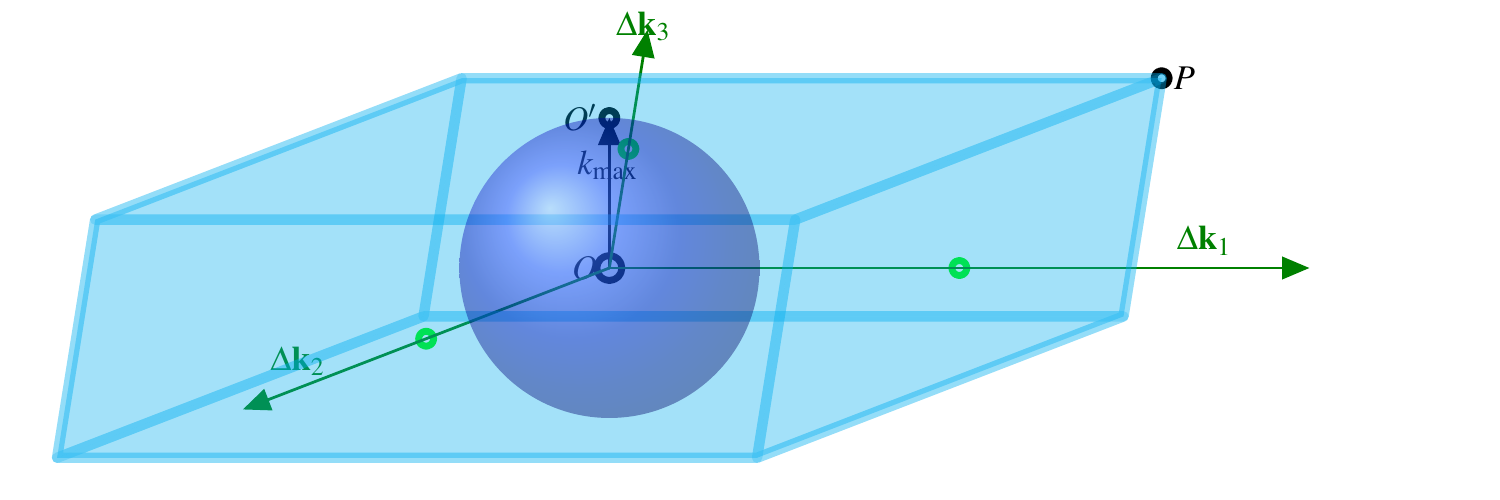}
  \end{center}
  \caption{%
  A sketch for the definition of the first-Brillouin-like zone,
  as formed with three vectors $\Delta\mathbf k_1$,
  $\Delta\mathbf k_2$, and $\Delta\mathbf k_3$.
  The zone is marked in transparent cyan.
  A sphere, defined the largest sphere
  contained in the zone, is rendered in blue.
  This figure is merely a sketch to show geometric
  relationships, instead of a plot using actual values.
  This figure shows the same concept as \citet{tjulin05}'s did.
  }\label{fig:def-sweetzone}
\end{figure}

In this sub-section, we propose the practical method
to estimate $\Delta \mathbf k$ by testing it with the virtual constellations.

Like the 1D case above, we introduce a monochromatic,
sinusoidal plane wave with a pre-assumed $\mathbf{k}$.
Then we use it to test our intended filters
(no matter whether they are Capon or MSR ones)
with a planned set of $\mathbf{k}'$.
If $\mathbf{k}'$ deviates from $\mathbf{k}$ greatly,
and at the same time the filter gives a $P$ large enough,
the aliasing effect can be regarded to happen.
This method is described as follows.
\begin{enumerate}
  \item Set a $\mathbf k$, an amplitude $A$,
    and thus make a Fourier-transformed image
    corresponding to the $\mathbf k$.
    Since the effects are decided by $\mathbf k - \mathbf{k}'$,
    the $\mathbf k$ can be chosen even as $\mathbf 0$.
    For simplicity, we actually do this in our example.
  \item Calculate all the $P_{\mathbf{k}'}$.
  \item Detect possible aliasing effects where
    $P \ge r A^2$ and
    $|\mathbf k' - \mathbf k| > \delta k$,
    where $r$ is an acceptable positive small threshold,
    and $\delta k$ is defined as the width of the filters.
\end{enumerate}

  As stated in the 1D case, if one chooses to use the filters directly without segment averaging,
  $r$ has to be small ($10^{-7}$ here) but
  large enough to distinguish the real effects
  from numerical ones introduced by $\epsilon$
  (accompanying the calculation of $\mathsf R^{-1}$).

As an illustration of our method,
in Figure~\ref{fig:solved-aliasing} we show the aliasing effects
by the constellations described in Sect.~2.3
and involving Capon filters only.
These diagrams show repeating patterns.
In sub-figure (a) and (d) (the 4-spacecraft constellation),
the undesired points are dense;
in (b) and (e) (5-spacecraft), they are sparser;
in (c) and (f) (9-spacecraft), they are very sporadic.
This comparison leads to an important conclusion:
an increase of the number of spacecraft in a constellation
will greatly enlarge the upper limit of $\mathbf k$
that a filtering method can measure,
thus giving a larger range of the first Brillouin zones.

We continue with a quantitative discussion
about the benefits of constellations with more spacecraft.
To record and analyse the dots in Figure~\ref{fig:solved-aliasing},
owing to the complexity of the repeating patterns in 3D,
we choose to record the location of the points in the $\mathbf k$-space
rather than to measure them directly on the diagram.
In the current version, we just find the 3 non-collinear
and non-coplanar modes with the smallest $|\mathbf k' - \mathbf k|$.
Here we take the case of the constellation with 4 spacecraft
(as in Figure~\ref{fig:solved-aliasing}, sub-plots (a) and (d))
as an example,
and justify our proposed quantitative calculation.
In the constellation, our method gives
\begin{linenomath*}
\begin{subequations}
  \begin{eqnarray}
    \Delta \mathbf k_1 &=& (-1/4, -1/4, 1/2), \\
    \Delta \mathbf k_2 &=& (0, -1/2, -1/2), \\
    \Delta \mathbf k_3 &=& (-3/4, 1/2, 0),
  \end{eqnarray}
  \label{eqn:solution4}
\end{subequations}
\end{linenomath*}
which restores our designed $\Delta \mathbf k$ expressed in Eq.~\eqref{eq:designed_k}, only with
some $(-1)$ factors on the vectors.
This indicates that our method has successfully
  calculated the $\Delta \mathbf k$
in the constellation with 4 spacecraft.

With the justification of our method above,
we are able to compare the first-Brillouin-like zones of our designed
constellations with different numbers of spacecraft.
With the same threshold of $P(\Delta \mathbf k)$ for
the aliasing effect, the zone can be computed.
The constellation with 5 spacecraft gives
\begin{linenomath*}
\begin{subequations}
  \begin{eqnarray}
    \Delta \mathbf k_1 &=& (-1/4, -3/4, 0), \\
    \Delta \mathbf k_2 &=& (-3/4, 0, -1/2), \\
    \Delta \mathbf k_3 &=& (-5/4, 1/2, 3/2);
  \end{eqnarray}
  \label{eqn:solution5}
\end{subequations}
\end{linenomath*}
the constellation with 9 spacecraft gives
\begin{linenomath*}
\begin{subequations}
  \begin{eqnarray}
    \Delta \mathbf k_1 &=& (-1, -3, 0), \\
    \Delta \mathbf k_2 &=& (-3, 0, -2), \\
    \Delta \mathbf k_3 &=& (-5, 2, 6).
  \end{eqnarray}
\end{subequations}
\end{linenomath*}

  In Figure~\ref{fig:compare-459}, we compare the parallelepiped
  spanned by $\Delta \mathbf k_1$, $\Delta \mathbf k_2$,
  and $\Delta \mathbf k_3$.
  In Figure~\ref{fig:compare-459}a, the benefits of the additional spacecraft are
  plotted comparatively.
  It illustrates the shape and dimensions of the $\Delta \mathbf k$ cells.
  Figure~\ref{fig:compare-459}b provides a close-up comparison between 4 and 5 spacecraft constellations.
  Here we see the improvement brought by the additional fifth spacecraft.
  We also compare the size of the cells with the spherical zones $k \le \pi / d_{\min{}}$,
  where $d_{\min{}}$ denotes the minimal distance between all the spacecraft in the constellation.
  In Figure~\ref{fig:compare-459}c, the cell is not even covered by the sphere.
  In Figure~\ref{fig:compare-459}d, the cell is enclosed within the sphere,
  but the sphere is obviously much larger than the cell.
  The radii of the spheres in (c) and (d)
  are 0.415 and 1.718, respectively,
  but from (b) we do not see such great difference.

To compare better the improvements by constellations with more spacecraft,
we need to compare typical parameters of Brillouin-like zones.
Here we choose to define the zones as \citet{tjulin05} did (nominated just as ``the parallelepiped'' therein),
i.e.~the parallelepiped defined by
\begin{linenomath*}
  \begin{equation}
    \Delta \mathbf k =
        c_1 \Delta \mathbf k_1
      + c_2 \Delta \mathbf k_2
      + c_3 \Delta \mathbf k_3,
      \qquad c_1, c_2, c_3 \in [-1/2, 1/2].
  \end{equation}
\end{linenomath*}
The zone is sketched as the cyan part in Figure~\ref{fig:def-sweetzone}.
From the parallelepiped, we can define a $k_{\max}$
as the radius of the largest sphere
that could be contained within the parallelepiped.
This radius can be computed as the minimum of $r$ calculated by
\begin{linenomath*}
  \begin{equation}
    r = \frac{|(\mathbf a \times \mathbf b)\cdot \mathbf c|}{2|\mathbf a \times \mathbf b|}
  \end{equation}
\end{linenomath*}
with $(\mathbf a, \mathbf b, \mathbf c)$ runs over all the permutations
of $(\Delta\mathbf k_1, \Delta\mathbf k_2, \Delta\mathbf k_3)$.
In this way, we calculate $k_{\max}$ for the constellations
with 4, 5, and 9 spacecraft correspondingly as
0.29, 0.38, and 1.52.
Here $k_{\max}$ increases greatly
as the number in the constellation increases.
Compared with the 4-spacecraft constellation,
the 5-spacecraft constellation yields a 30\% improvement
of $k_{\max}$ with only a 25\% larger number of spacecraft.
The 9-spacecraft counterpart, also compared to the 4-spacecraft one,
has a $k_{\max}$ of 5.19 times with only a 2.25 times number of spacecraft.
  Also, we calculate the volumes of the cells.
  For the 4, 5, and 9-spacecraft constellations,
  the volumes are respectively 0.344, 1.375, and 88.0.

\subsection{Application of the Practical Method to the MMS Constellation}

For a more realistic case, we choose an MMS case,
i.e.~the MMS constellation at 16:23:24 UT on 1 December 2016.
At that point, the four spacecraft were located at
$\mathbf x_1 = (-0.324, -5.578, -2.304)\;\textrm{km}$,
$\mathbf x_2 = (-3.833, -2.337,  2.591)\;\textrm{km}$,
$\mathbf x_3 = ( 3.940, -3.830,  2.062)\;\textrm{km}$,
$\mathbf x_4 = (-1.136, -8.166,  2.722)\;\textrm{km}$, respectively.
The values are in GSE coordinates,
but subtracted with an offset $(54630, 43600, 7690)\;\textrm{km}$.
From the distances between the spacecraft,
a range of $\Delta k$ could be estimated within 2~km$^{-1}$.
With a progressively finer series of grids at resolutions from $0.05\;\textrm{km}^{-1}$ (initial)
to $0.001\;\textrm{km}^{-1}$ (final),
we get $\Delta \mathbf k_1 = (0.917, 0.472, 0.371)\;\textrm{km}^{-1}$,
$\Delta \mathbf k_2 = (0.636, -0.791, -0.304)\;\textrm{km}^{-1}$,
and $\Delta \mathbf k_3 = (0.813, 0.355, -0.936)\;\textrm{km}^{-1}$
(for details in these computations, please see Dataset S1 available at \citet{dataset_S1_S2}).
Since there are only 4 spacecraft, the exact solutions can be solved by \citet{neubauer90}'s and \citet{tjulin05}'s method,
and gives a cell with three non-coplanar $\Delta \mathbf k$
with smallest $|\Delta \mathbf k|$ as
$(0.636, -0.791, -0.304)\;\textrm{km}^{-1}$,
$(0.917, 0.432, 0.371)\;\textrm{km}^{-1}$,
and $(0.813, 0.355, -0.936)\;\textrm{km}^{-1}$.
Our method gives the occurrences approximated to 0.001~km$^{-1}$,
without any previous knowledge of the occurrences.

\section{Reconstruction of 4D Power Spectra of Turbulence-like Data within the Identified First-Brillouin-like Zones}

Here we present our attempts to reconstruct 4D turbulence spectra.
  We choose the ``slab + 2D'' model to investigate the
  $\mathbf k$-filtering technique with constellations formed by different numbers of spacecraft.
  From a viewpoint of technique verification,
  the ``slab'' part gives a benchmark test for
  a one-to-one correspondence between $\omega$ and $\mathbf k$,
  and the ``2D'' part represents a general case where there are
  more modes ($\mathbf k$'s) for a single $\omega$.
  In the present work, the synthetic fluctuation fields are generated from the ``slab + 2D'' model,
  based on which the field sampling is conducted
  by multiple virtual spacecraft
  and the 3D PSD is reconstructed by employing the $\mathbf k$-filtering technique.
  In the future, we plan to apply this technique
  to reconstruct the 3D PSD of turbulence
  as simulated in 3D models and measured by multiple virtual
  spacecraft.

\subsection{Settings of Signal Generator and Filters}

We utilise the conditions of aliasing effects above
to conduct comparison tests.
It is easier to start with the ``slab only'' scenario
because of its simplicity,
and then we design the ``slab + 2D'' test.

\subsubsection{Choices of $\mathbf k$ and $\mathbf k'$}

With our Brillouin zones in Sect.~3, we design a uniformly
distanced grid in the wavevector space with a stencil distance $\delta k = 1/16$.
This stencil distance applies both to $\mathbf k'$ and to $\mathbf k$.
Therefore, the signal generator uses $\mathbf k$
as the list $(1/16, 1/16, 1/16)$, $(1/8, 1/8, 1/8)$,
$(-1/16, -1/16, -1/16)$, and $(-1/8, -1/8, -1/8)$.
As to the $\mathbf k'$, we make a grid $(p/16, q/16, r/16)$,
but except for $\mathbf k'= (0, 0, 0)$.
In this way, the aliasing effect for $\mathbf k = (1/8, 1/8, 1/8)$
occurs at $\mathbf k' = (1/8, -3/8, -3/8)$.

  The range of $p$, $q$, and $r$ can be chosen more freely.
  In the ``slab only'' case, a wider range $\mathbf k'$ is set,
  so that the simpler case can illustrate the pattern of aliasing effects.
  In the ``slab + 2D'' case, since the wave modes are much more,
  the range of $\mathbf k'$ is chosen smaller so that clearer plots are made.

For the ``slab + 2D'' test, for the sake of simplicity,
we use the same settings of $\mathbf k$ for the slab part,
and the addition of the 2D part is designed as
\begin{linenomath*}
  \begin{equation}
    \mathbf k = (p, q, r) \delta k, \text{ with } p + q + r = 0 \text{ and } \mathbf k \ne \mathbf 0,
    \label{eqn:kpqr}
  \end{equation}
\end{linenomath*}
where $p$, $q$, and $r$ are integers.
We just take the upper limit of $|p|, |q|, |r|$ as 5 in both cases,
and keep the modes with $\sqrt{p^2 + q^2 + r^2} \le 6$.
Nevertheless, it is also possible to scheme the tests
and to design perfectly conditions
so that the aliasing effects as expected.

\subsubsection{Signal Generator}

For the slab part, we generate the modes according to
the $\mathbf k$ designed above.
It is rather easy to generate the slab part by assigning
\begin{linenomath*}
  \begin{equation}
    |\delta \mathbf B(\mathbf k)| = C \cdot
      \left(\frac{|\mathbf k|}{k_{\min}}\right)^{-1},
  \end{equation}
\end{linenomath*}
where $C$ is a pre-assumed amplitude, and $k_{\min}$ denotes the magnitude of the smallest
wave-vector of all the slab modes.

For the 2D part, the amplitudes and the modes need
more complicated arrangements.
A wave vector of a given 2D mode in Eq.~\eqref{eqn:kpqr} requires
$p + q + r = 0$.
Meanwhile, the spectral index of the integrated / reduced spectrum is set so that $\text{PSD}(k) \propto k^{-5/3}$.
The total energy of the 2D modes is set to be 5 times of that of the slab modes.
The modes are tabulated in Dataset S2 available at \citet{dataset_S1_S2}.

The next issue is $\omega$ and consequently
the temporal resolution of sampling $\Delta t$.
From the dispersion relation (see Sect.~2.2),
for a mode with the wave vector $(p, q, r)\delta k$,
the corresponding angular frequency is
\begin{linenomath*}
  \begin{equation}
    \omega = \delta k v_{\mathrm A} \cdot (p + q + r) / \sqrt{3} + \delta k v_{\textrm{SW}} n
    = \frac{\delta k \cdot (p + q + r + 21r/2)}{\sqrt3}
  \end{equation}
\end{linenomath*}
For the slab modes, $p = q = r$, and $\omega = 9\sqrt3\delta k r/2$;
for the 2D modes, $p + q + r = 0$, and $\omega = 7\sqrt3\delta kr/2$.
Therefore, the values of $\omega$ are aligned on a stencil
whose resolution is $\delta\omega=\sqrt3\delta k/2$.
Since all the $\mathbf k$'s corresponding to the 2D modes
share the same 3D grid with the slab modes,
and the slab modes have a larger $\omega$ with $r$,
it suffices to plan with the maximal $r$.
The maximal $r$ is 3, so
$\omega_{\max} = 27\sqrt3\delta k/2 = 27\sqrt3/32$, or $27\delta\omega$.
Here we use $27 \times 4 = 108$ samples in a time series
to resolve the signal, and accordingly
$\Delta t = (2\pi / \delta\omega) / 108 = 16\pi/(27\sqrt3)
\approx 1.075$.

In the implementations, we also made the following arrangements.
Firstly, to provide enough number of degrees of freedom and
to make $\mathsf{R}$ more statistically significant \citep[e.g.][]{narita11},
$\mathsf{R}$ must be averages of FFTs of many (24 here) segments of $\mathbf{B}_{ij}$ (see Sect.~2.1).
Secondly, for the 2D modes, akin to \citet{narita11}'s test case,
we add a slight random shift of $\lesssim 0.1\%$ on $\omega_j$.

\subsection{Results of the ``Slab Only'' Case}

\begin{figure}[t]
  \begin{center}
    \includegraphics[width=\textwidth]{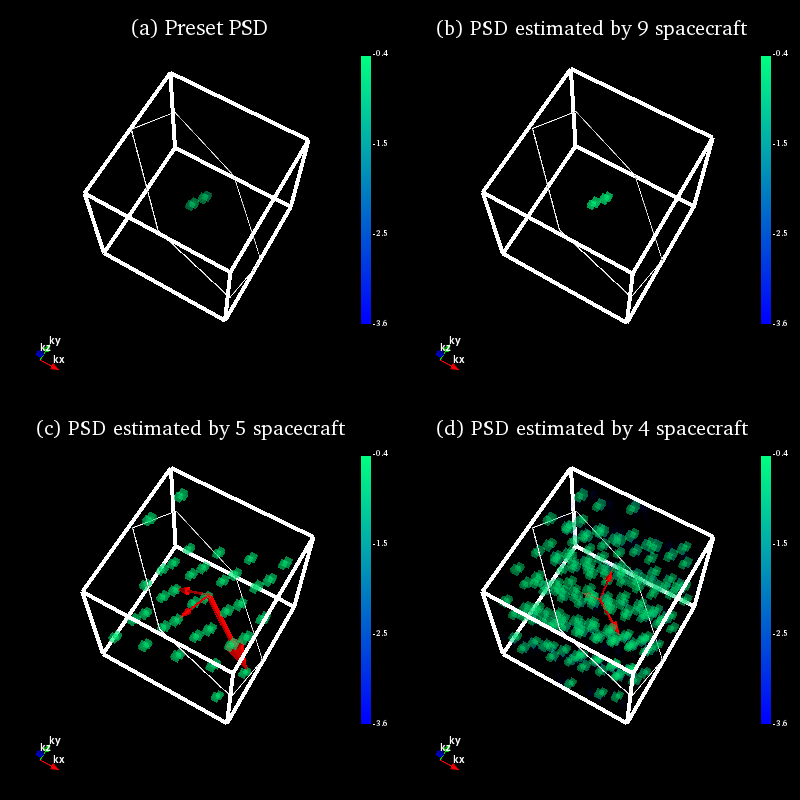}
  \end{center}
  \caption{%
    $\textrm{PSD}(\mathbf k)$ for the ``slab only'' case,
    consisting of (a) the preset PSD
    and PSD reconstructed by the constellations
    with (b) 9, (c) 5, and (d) 4 virtual spacecraft.
    Colours represent logarithms of PSD values.
    For reference, the ``2D'' plane (perpendicular to $\mathbf B_0$)
    is also plotted in thin white frames.
    In (c) and (d), the vectors $\Delta \mathbf k_1$ to $\Delta \mathbf k_3$
    are plotted with red arrows.
  }\label{fig:PSD_test1_slabonly}
\end{figure}

We start with the simpler ``slab only'' case and compare
results by the three constellations.
To compare the difference in aliasing effects between the constellations,
in Figure~\ref{fig:PSD_test1_slabonly},
we plot $\textrm{PSD}(\mathbf k)$ of the ``slab only'' wave packet.
For reconstructed PSD, the quantity $\textrm{PSD}(\mathbf k)$
is calculated as summing over all $\omega$.
Since the test signal is seriously designed for $\omega$,
and our aim of the tests is not concentrated in temporal effects,
though we have conducted 4D PSD reconstructions,
we choose to omit the dimension $\omega$.
As to $\omega$, the modes occur at the preset parts.
Here, just as in Figure~\ref{fig:solved-aliasing},
the $\mathbf k$-points lit-up in green denote
the wave vectors corresponding to large energy.

In order to recognize the aliasing effects,
note that the preset modes are only four along the slab direction (Figure \ref{fig:PSD_test1_slabonly}a).
We can compare between the preset diagram and the reconstructed ones
to identify the aliasing effects.
The reconstructed PSD as derived from the measurements by 9 virtual spacecraft
(see Figure \ref{fig:PSD_test1_slabonly}b) looks almost identical to the preset PSD
in Figure \ref{fig:PSD_test1_slabonly}a.
and no occurrences of aliasing effects are detected,
while the PSD of the ``slab only'' wave packet as estimated by 5 and 4 virtual spacecraft
apears periodically in the box of wavevector space
(see Figure \ref{fig:PSD_test1_slabonly}c and \ref{fig:PSD_test1_slabonly}d, respectively).
  To quantitatively illustrate the occurrences of aliasing effects,
  the vectors $\Delta \mathbf k_1$ to $\Delta \mathbf k_3$
  are supplied as the red arrows in Figure \ref{fig:PSD_test1_slabonly}c and \ref{fig:PSD_test1_slabonly}d.

Hence, the aliasing effect greatly influences the results,
just as designed.
Therefore, the results testify our method
to compute for aliasing effects.
  Also, it is noteworthy that the 9-spacecraft constellation
  does not eliminate aliasing effects.
  More spacecraft can alleviate aliasing effects with more sampled data
  and scarcer occurrences of aliasing effects,
  but aliasing effects can never be eliminated.
  Such effects are not present in Figure \ref{fig:PSD_test1_slabonly}b,
  only because of their large $\Delta \mathbf k$.

\subsection{The ``Slab+2D'' Case}

\begin{figure}[t]
  \begin{center}
    \includegraphics[width=\textwidth]{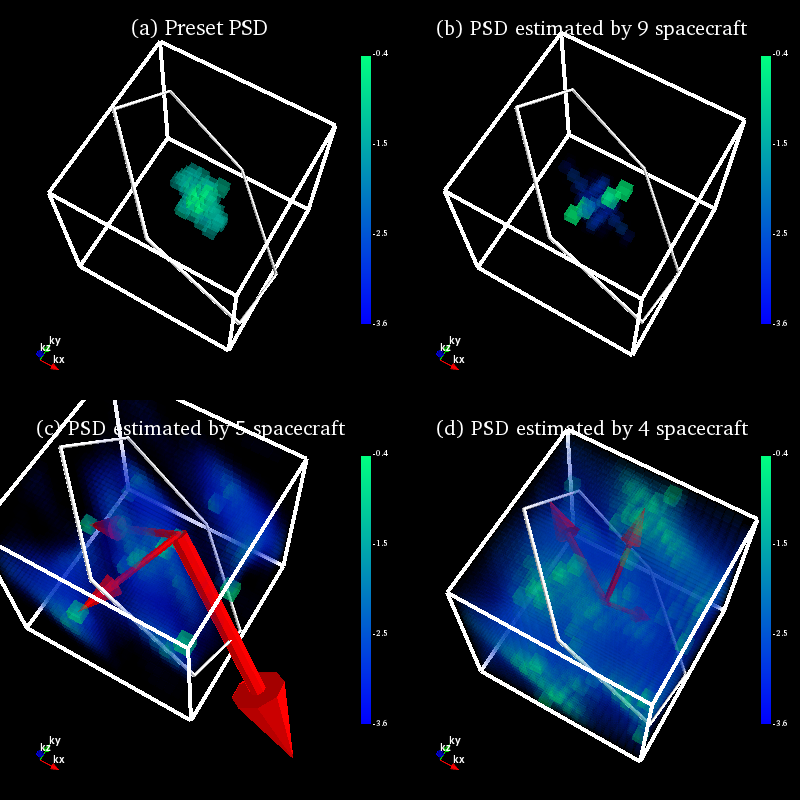}
  \end{center}
  \caption{$\textrm{PSD}(\mathbf k)$ in the ``slab + 2D'' case.
    The format is similar to Figure~\ref{fig:PSD_test1_slabonly}'s.
  }%
  \label{fig:PSD_test1_slab2D}
\end{figure}

This case is more similar to actual solar wind turbulence
in the ``slab+2D'' interpretation than the previous case.
Here the preset signals are more complicated,
and there are more
wave modes for a single $\omega$ than in the ``slab only'' case.
In this part, we report the benchmark test in a manner similar to the
``slab only'' case.

To check the aliasing effects, one first compares
Figure \ref{fig:PSD_test1_slab2D}a to \ref{fig:PSD_test1_slab2D}d,
where greenish repeated patterns are to be noticed
rather than the blueish mists.
One sees again the patterns recur, just as in the
``slab only'' case.
The only difference of the patterns is merely visual,
where in the ``slab only'' case,
the occurrences of aliasing effects are clearly seen,
and here in the ``slab + 2D'' case,
some occurrences are blocked from sight because of the blueish mist.
The comparison between \ref{fig:PSD_test1_slab2D}a and \ref{fig:PSD_test1_slab2D}c also displays
aliasing effects obtained from the 5-spacecraft constellation,
as in the ``slab only'' case.
The most convincing comparison among the three results
is between \ref{fig:PSD_test1_slab2D}a and \ref{fig:PSD_test1_slab2D}b,
where no repeated patterns are detected.
In terms of aliasing effects, this test also testifies
the computations in Section 3.2.

\section{Summary and Discussion}

\begin{figure}[t]
  \begin{center}
    \includegraphics[width=\textwidth]{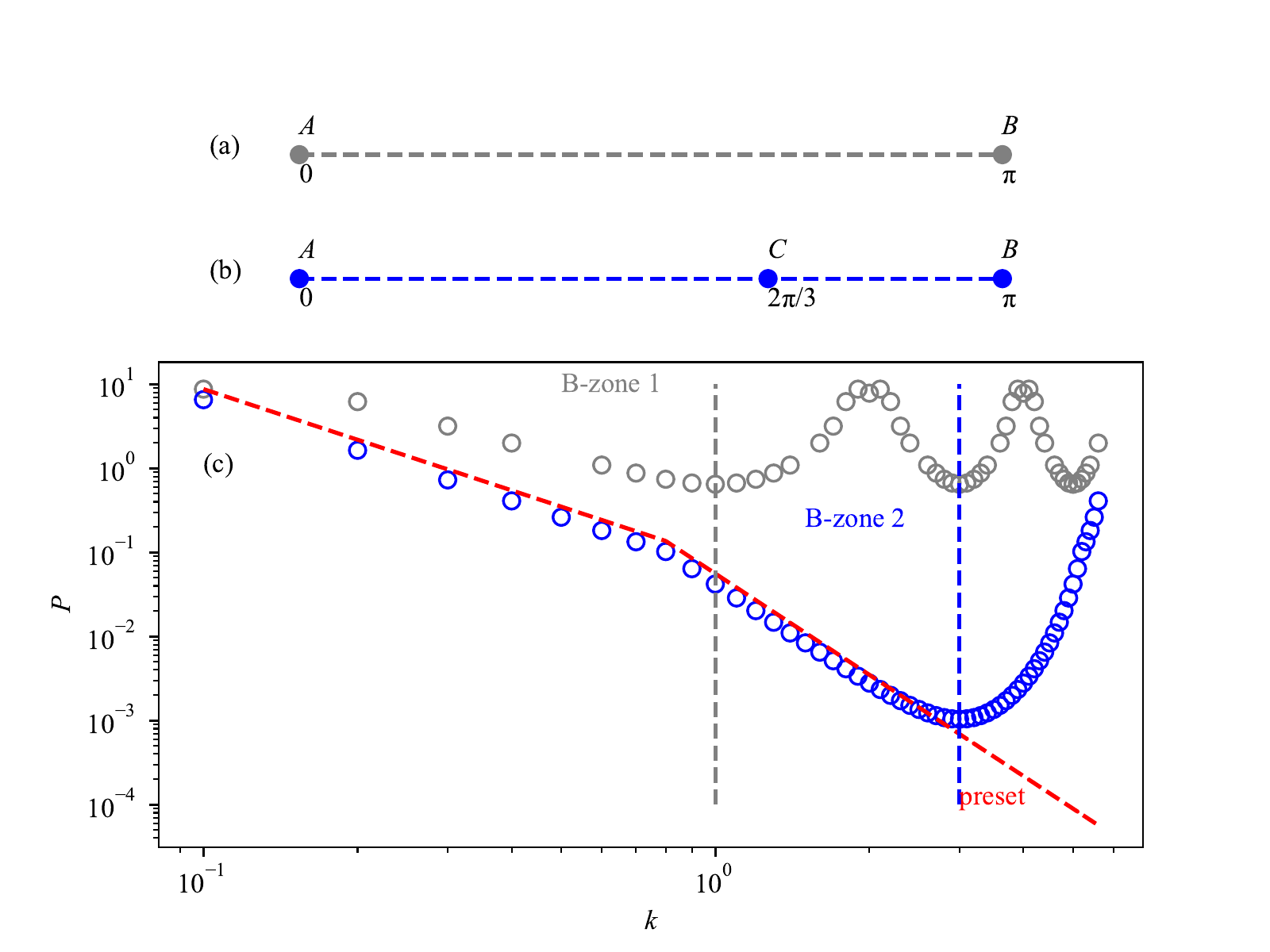}
  \end{center}
  \caption{
  \begin{rev}
    (a \& b)Two constellation configurations with different separations.
    (c)The preset PSD (red dashed line) and reconstructed
    PSDs (blue circles for the configuration in (b), grey circles for the configuration in (a)).
  \end{rev}
  }%
  \label{fig:breaking_point}
\end{figure}

In the context of complex constellation configurations,
the aliasing effect also becomes complicated.
To compute such effects, we have proposed a practical
method
  to calculate the $\mathbf k'$ where aliasing effects occur
in this article.
The method works by simulating aliasing effects of
a monochromatic sinusoidal plane wave with the same multi-channel
analysis technique as the filtering method employed by actual 4D spectrum
reconstruction,
and such results can be tested with more realistic signals.
This method brings about some attempts to explore
how different constellation configurations can help constrain the aliasing effects.
This method can lead to a better understanding of
aliasing effects imposed by multi-spacecraft constellations,
especially under the cross-scale circumstances,
\begin{rev}
with constellations of more than four spacecraft,
e.g. HelioSwarm \citep{klein19} or Prospero \citep{retino19}.
\end{rev}

As mentioned above, the estimation of $\Delta\mathbf k$ for the first-Brillouin-like zone
can be verified practically.
The practical verification is displayed by the benchmark
tests and the test of actual MMS positions in this article.
  Meanwhile, we also discuss some 1D cases for an analogy to the 3D cases,
  but the 1D model itself might be applicable to suggested exploration proposals
  as the Debye mission proposal \citep{verscharen19, debyemission}.
  In 1D cases, a smaller distance $d'$ yields to a larger Brillouin-like zone,
  but the $d'$ should be interpreted as a greatest-common-divisor-like grid distance
  (as if the spacecraft are arranged on a virtual grid),
  not a geometric distance, and in some cases the $d'$ is hard to be defined strictly.
  An estimation $\textrm{PSD}(k')$ of a sinusoidal wave can identify the occurrences of
  aliasing effects practically.
  In 3D cases, the additional spacecraft bring larger Brillouin-like zones,
  because they are added inside the basic tetrahedron and aim to observe smaller-scale oscillations.

\begin{rev}
Also,
to emphasize the advantage of a wider first Brillouin zone,
we consider a set of 1D cases with a turning point
between two power-law-like segments of Alfv\'en waves
(like the ``slab only'' case in Sect. 4),
similar to the turning points in solar wind turbulence
(see the details in Figure \ref{fig:breaking_point}c).
The configurations of constellations are shown respectively in
Figures \ref{fig:breaking_point}a and \ref{fig:breaking_point}b.
In Case 1, $d' = \pi$ so that the first Brillouin zone is $k \in [-1, 1]$;
in Case 2, $d' = \pi/3$, so the first Brillouin zone is $k \in [-3, 3]$.
The corresponding first Brillouin zones are coloured in gray and blue in Figure \ref{fig:breaking_point}c.
The reconstructed PSDs are plotted in the same colors, respectively.
The bigger $d'$ in Case 1 brings a narrower first Brillouin zone
(equivalently, the signals are not sampled well),
and the turning point is not resolved.
In Case 2, the smaller $d'$ brings a wider first Brillouin zone,
which enables a better reconstruction of PSD near the turning point.
\end{rev}

However, there are several issues before this method
is employed in actual spacecraft configuration planning
or data process.
Firstly, though the method is practical,
the interpretations need further investigation.
In this article, we discuss the aliasing effect
by three non-planar ``smallest'' $\Delta \mathbf k$,
forming a parallelepiped cell.
This might be an over-simplification,
since in cross-scale constellations, more complicated
patterns of aliasing effects will be shown in the wave-vector
space than repeating parallelepiped cells.
A more general geometric description of the Brillouin-like zones is required.
Owing to their inhomogeneous feature, the finer parameters
might be difficult to extract from the estimation,
and $k_{\max{}}$ in Figure~\ref{fig:def-sweetzone} might be an
underestimation for the available range of $\mathbf k$.

In actual observations, some effects must be considered.
Fortunately, the remedies have mostly been well known.
The motions of the spacecraft should not be neglected as what we have done here.
Since this mainly involves the Doppler shift,
a trivial transform to spacecraft frame is needed.
If one must consider the effect when each spacecraft orbits
with its own velocity, a careful re-modelling should be conducted.
Some filtering methods, like the one presented herein,
do not function well if oscillations along a certain axis
($x$, $y$, or $z$) are small
\begin{rev}
(which applies both to components of $\mathbf k$ and
to components of vector fields),
\end{rev}
owing to the numerical methods employed.
Its remedy is simply a rotation of coordinates system: for example, in the case when oscillations along $z$ are small,
one can just take $\xi = x$, $\eta = (y + z) / \sqrt 2$, and
$\zeta = (y - z) / \sqrt 2$,
and then apply the filtering method in the $(\xi, \eta, \zeta)$
system.

Some possible future work can be done to extend this method.
Such filtering method is actually suitable for parallel computing,
because the computational load is easily and evenly divided along $\omega$
or $\mathbf k'$, and little dependence occurs among the divided parts.
\begin{rev}
The filtering methods could also be improved with more physics considered,
e.g. the inclusion of the Maxwell-Ampere equation
$\mathbf j = \nabla \times \mathbf B = \rho\cdot(\mathbf v_{\textrm i} - \mathbf v_{\textrm e})$. 
\end{rev}
It might be computationally practical to yield high-resolution 4D PSD
if the filtering method is completed,
because the heavy computation can be parallelized.
Moreover, some algorithms might be similarly developped
to model certain features as discontinuities, cross-scale pressure balance structures,
magnetic flux tubes etc., using cross-scale observations.
\begin{rev}
We would feel honoured if we can work together with you on such a challenging problem.
We are also thinking about how to get the 4D phase spectra with some techniques
like the 4D power spectra as the k-filtering method already provides us.
If the 4D phase spectra could be derived from the data,
then the 4D spectra of dissipation/conversion rate $\mathbf j \cdot \mathbf E$
and the Poynting vector $\mathbf E \times \mathbf B$
are anticipated to be yielded.
Still, We learned that machine learning had been successfully applied
to solve inversion problems in geophysics.
We hope that the amplitude and phase spectra may also be derived
from the measurements
when the machine learning scheme is well established and well trained
with massive modelling data.
\end{rev}

\acknowledgments\
This research is supported by the NSFC under contracts 41574168, 41231069,
41274172, 41474148, 41421003, 41574166, 41531073,
41731067, and 41804164.
LZ is also supported by the 13th Five-year Informatisation Plan
of Chinese Academy of Sciences, Grant No.~XXH13505-04.
The authors are grateful for valuable discussions with Prof.~C.Y.~Tu and
Prof.~E.~Marsch.
The authors are also grateful for Prof.~Dr.~Sahraoui's suggestions
as a reviewer, 
and his idea of introducing more physics to imporve the filtering methods.
The MMS data used in this work are available from the MMS Science Data Center (\url{http://lasp.colorado.edu/mms/sdc}).
Datasets S1 and S2 for this research are available in the in-text data citation references
\citep{dataset_S1_S2}. 
This research is conducted with help of open-source software
  Eigen \citep*{eigenweb}, %
  FFTW \citep{fftw05}, %
  matplotlib \citep{matplotlib}, %
  and Mayavi \citep{mayavi};
the plotting of the curly braces in Figure~3 is helped from Dr.~Siyu Gao's
\texttt{matplotlib-curly-brace} package available at
\url{https://github.com/iruletheworld/matplotlib-curly-brace}.

\bibliographystyle{agufull08}

\begin{thebibliography}{51}
\providecommand{\natexlab}[1]{#1}
\expandafter\ifx\csname urlstyle\endcsname\relax
  \providecommand{\doi}[1]{doi:\discretionary{}{}{}#1}\else
  \providecommand{\doi}{doi:\discretionary{}{}{}\begingroup
  \urlstyle{rm}\Url}\fi

\bibitem[{\textit{{Balogh} et~al.}(1997)\textit{{Balogh}, {Dunlop}, {Cowley},
  {Southwood}, {Thomlinson}, {Glassmeier}, {Musmann}, {Luhr}, {Buchert},
  {Acuna}, {Fairfield}, {Southwoodlavin}, {Riedler}, {Schwingenschuh}, and
  {Kivelson}}}]{balogh97}
{Balogh}, A., M.~W. {Dunlop}, S.~W.~H. {Cowley}, D.~J. {Southwood}, J.~G.
  {Thomlinson}, K.~H. {Glassmeier}, G.~{Musmann}, H.~{Luhr}, S.~{Buchert},
  M.~H. {Acuna}, D.~H. {Fairfield}, J.~A. {Southwoodlavin}, W.~{Riedler},
  K.~{Schwingenschuh}, and M.~G. {Kivelson} (1997), {The Cluster Magnetic Field
  Investigation}, \textit{Space Sci. Rev.}, \textit{79}, 65--91,
  \doi{10.1023/A:1004970907748}.

\bibitem[{\textit{{Bieber} et~al.}(1996)\textit{{Bieber}, {Wanner}, and
  {Matthaeus}}}]{bieber96}
{Bieber}, J.~W., W.~{Wanner}, and W.~H. {Matthaeus} (1996), {Dominant 2D
  magnetic turbulence in the solar wind}, in \textit{American Institute of
  Physics Conference Series}, \textit{American Institute of Physics Conference
  Series}, vol. 382, edited by D.~{Winterhalter}, J.~T. {Gosling}, S.~R.
  {Habbal}, W.~S. {Kurth}, and M.~{Neugebauer}, pp. 355--357,
  \doi{10.1063/1.51410}.

\bibitem[{\textit{Boldyrev}}(2006)\textit{Boldyrev}]{boldyrev06}
{Boldyrev}, S. (2006), {Spectrum of Magnetohydrodynamic Turbulence},
  \textit{{Phys. Rev. Lett.}}, \textit{96}, 115002.

\bibitem[{\textit{Brandenburg and Lazarian}(2013)}]{brandenburg13}
Brandenburg, A., and A.~Lazarian (2013), Astrophysical hydromagnetic
  turbulence, \textit{{Space Sci. Rev.}}, \textit{178}, 163--200,
  \doi{10.1007/s11214-013-0009-3}.

\bibitem[{\textit{{Bruno} and {Carbone}}(2013)}]{bruno13}
{Bruno}, R., and V.~{Carbone} (2013), {The Solar Wind as a Turbulence
  Laboratory}, \textit{Living Reviews in Solar Physics}, \textit{10}, 2.

\bibitem[{\textit{{Burch} et al.}(2016)\textit{{Burch}, {Moore}, {Torbert}, and {Giles}}}]{burch16}
{Burch}, J.~L., T.~E. {Moore}, R.~B. {Torbert} and B.~L. {Giles} (2016),
{Magnetospheric Multiscale Overview and Science Objectives},
  \textit{Space Sci. Rev.},
  \textit{199}, 5--21.

\bibitem[{\textit{{Burlaga}}(1968)}]{burlaga68}
{Burlaga}, L. (1968), {Micro-scale Structures in the Interplanetary Medium},
  \textit{Solar Physics}, \textit{4}, 67--92.

\bibitem[{\textit{Burlaga}(1971)}]{burlaga71}
Burlaga, L. (1971), {Hydromagnetic waves and discontinuities in the solar
  wind}, \textit{Space Science Reviews}, \textit{12}, \doi{10.1007/BF00173345}.

\bibitem[{\textit{{Capon}}(1969{\natexlab{a}})}]{capon69}
{Capon}, J. (1969{\natexlab{a}}), {High Resolution Frequency-Wavenumber
  Spectrum Analysis}, in \textit{Proc. IEEE, Volume 57, p. 1408-1418}, vol.~57,
  pp. 1408--1418.

\bibitem[{\textit{{Capon}}(1969{\natexlab{b}})}]{capon69geoph}
{Capon}, J. (1969{\natexlab{b}}), {Long-Period Signal Processing Results for
  the Large Aperture Seismic Array}, \textit{Geophysics}, \textit{34}, 305,
  \doi{10.1190/1.1440014}.

\bibitem[{\textit{{Capon}}(1969{\natexlab{c}})}]{capon69jgr}
{Capon}, J. (1969{\natexlab{c}}), {Investigation of long-period noise at the
  large aperture seismic array}, \textit{J. Geophys. Res.}, \textit{74},
  3182--3194, \doi{10.1029/JB074i012p03182}.

\bibitem[{\textit{{Chanteur} and {Roux}}(1993)}]{chanteur93}
{Chanteur}, G., and A.~{Roux} (1993), {A European network for the numerical
  simulation of space plasmas}, in \textit{Cluster: Mission, Payload and
  Supporting Activities}, \textit{ESA Special Publication}, vol. 1159, edited
  by W.~R. {Burke}, p. 307.

\bibitem[{\textit{{Chen} et~al.}(2010)\textit{{Chen}, {Horbury},
  {Schekochihin}, {Wicks}, {Alexandrova}, and {Mitchell}}}]{chen10}
{Chen}, C.~H.~K., T.~S. {Horbury}, A.~A. {Schekochihin}, R.~T. {Wicks},
  O.~{Alexandrova}, and J.~{Mitchell} (2010), {Anisotropy of Solar Wind
  Turbulence between Ion and Electron Scales}, \textit{Phys. Rev. Lett.},
  \textit{104}(25), 255002, \doi{10.1103/PhysRevLett.104.255002}.

\bibitem[{\textit{{Chen}}(2016)}]{chen16}
{Chen}, C.~H.~K. (2016),
  {Recent progress in astrophysical plasma turbulence from solar wind observations},
  \textit{J. Plasma Phys.}, \textit{82}, 535820602.

\bibitem[{\textit{Cho and Lazarian}(2003)}]{cho03}
Cho, J., and A.~Lazarian (2003), {Compressible magnetohydrodynamic turbulence:
  mode coupling, scaling relations, anisotropy, viscosity-damped regime and
  astrophysical implications}, \textit{{Mon. Not. R. Astron. Soc.}},
  \textit{345}, 325--339, \doi{10.1046/j.1365-8711.2003.06941.x}.

\bibitem[{\textit{Cho and Vishniac}(2000)}]{cho00}
Cho, J., and E.~Vishniac (2000), The anisotropy of magnetohydrodynamic alfvenic
  turbulence, \textit{{ApJ}}, \textit{539}(1), 273,282, \doi{10.1086/309213}.

\bibitem[{\textit{Cho et~al.}(2002)\textit{Cho, Lazarian, and
  Vishniac}}]{cho02}
Cho, J., A.~Lazarian, and E.~T. Vishniac (2002), {Simulations of
  Magnetohydrodynamic Turbulence in a Strongly Magnetized Medium},
  \textit{{Astrophys. J.}}, \textit{564}, \doi{10.1086/324186}.

\bibitem[{\textit{Choi et~al.}(1993)\textit{Choi, Song, and Kim}}]{choi93}
Choi, J., I.~Song, and H.~M. Kim (1993), On estimating the direction of arrival
  when the number of signal sources is unknown, \textit{Signal Processing},
  \textit{34}(2), 193 -- 205,
  \doi{http://dx.doi.org/10.1016/0165-1684(93)90162-4}.

\bibitem[{\textit{{Dunlop} et~al.}(1990)\textit{{Dunlop}, {Balogh},
  {Southwood}, {Elphic}, {Glassmeier}, and {Neubauer}}}]{dunlop90}
{Dunlop}, M.~W., A.~{Balogh}, D.~J. {Southwood}, R.~C. {Elphic}, K.-H.
  {Glassmeier}, and F.~M. {Neubauer} (1990), {Configurational sensitivity of
  multipoint magnetic field measurements}, \textit{Tech. rep.}

\bibitem[{\textit{{Escoubet} et~al.}(1997)\textit{{Escoubet}, {Schmidt}, and
  {Goldstein}}}]{escoubet97}
{Escoubet}, C.~P., R.~{Schmidt}, and M.~L. {Goldstein} (1997), {Cluster -
  Science and Mission Overview}, \textit{Space Sci. Rev.}, \textit{79}, 11--32,
  \doi{10.1023/A:1004923124586}.

\bibitem[{\textit{{Escoubet} et~al.}(2001)\textit{{Escoubet}, {Fehringer}, and {Godlstein}}}]{escoubet01}
{Escoubet}, C.~P., M.~{Fehringer}, and M.~{Goldstein} (2001),
{Introduction The Cluster mission}, \textit{Ann. Geophys.},
  \textit{19}, 1197--1200.

\bibitem[{\textit{{Forman} et~al.}(2011)\textit{{Forman}, {Wicks}, and
  {Horbury}}}]{forman11}
{Forman}, M.~A., R.~T. {Wicks}, and T.~S. {Horbury} (2011), {Detailed Fit of
  ``Critical Balance'' Theory to Solar Wind Turbulence Measurements},
  \textit{Astrophys. J.}, \textit{733}, 76, \doi{10.1088/0004-637X/733/2/76}.

\bibitem[{\textit{Frigo and Johnson}(2005)}]{fftw05}
Frigo, M., and S.~G. Johnson (2005), {The Design and Implementation of
  {FFTW3}}, \textit{Proceedings of the IEEE}, \textit{93}(2), 216--231, special
  issue on ``Program Generation, Optimization, and Platform Adaptation''.

\bibitem[{\textit{Glassmeier et al.}(2001)}]{glassmeier01}
Glassmeier, K.-H., U.~Motschmann, M.~Dunlop, A.~Balogh,
M.H.~{Acu\~na}, C.~Carr, G.~Musmann, K.H.~Fornacon, K.~Schweda,
J.~Vogt, G.~Georgescu, and S.~Buchert (2001),
  Cluster as a wave telescope -- first results from the fluxgate magnetometer
  \textit{Ann. Geophys.}, \textit{19}, 1439--1447.

\bibitem[{\textit{{Goldreich} and {Sridhar}}(1995)}]{goldreich95}
{Goldreich}, P., and S.~{Sridhar} (1995), {Toward a theory of interstellar
  turbulence. 2: Strong alfvenic turbulence}, \textit{{Astrophys. J.}},
  \textit{438}, 763--775, \doi{10.1086/175121}.

\bibitem[{\textit{Guennebaud et~al.}(2010)\textit{Guennebaud, Jacob
  et~al.}}]{eigenweb}
Guennebaud, G., B.~Jacob, et~al. (2010), Eigen v3, http://eigen.tuxfamily.org.

\bibitem[{\textit{{He} et~al.}(2011)\textit{{He}, {Marsch}, {Tu}, {Yao}, and
  {Tian}}}]{he11}
{He}, J., E.~{Marsch}, C.~{Tu}, S.~{Yao}, and H.~{Tian} (2011), {Possible
  Evidence of Alfv{\'e}n-cyclotron Waves in the Angle Distribution of Magnetic
  Helicity of Solar Wind Turbulence}, \textit{{Astrophys. J.}}, \textit{731},
  85, \doi{10.1088/0004-637X/731/2/85}.

\bibitem[{\textit{{He} et~al.}(2012)\textit{{He}, {Tu}, {Marsch}, and
  {Yao}}}]{he12}
{He}, J., C.~{Tu}, E.~{Marsch}, and S.~{Yao} (2012), {Do Oblique
  Alfv{\'e}n/Ion-cyclotron or Fast-mode/Whistler Waves Dominate the Dissipation
  of Solar Wind Turbulence near the Proton Inertial Length?},
  \textit{{Astrophys. J. Letters}}, \textit{745}, L8,
  \doi{10.1088/2041-8205/745/1/L8}.

\bibitem[{\textit{{He} et~al.}(2013)\textit{{He}, {Tu}, {Marsch}, {Bourouaine},
  and {Pei}}}]{he13}
{He}, J., C.~{Tu}, E.~{Marsch}, S.~{Bourouaine}, and Z.~{Pei} (2013), {Radial
  Evolution of the Wavevector Anisotropy of Solar Wind Turbulence between 0.3
  and 1 AU}, \textit{Astrophys. J.}, \textit{773}, 72,
  \doi{10.1088/0004-637X/773/1/72}.

\bibitem[{\textit{Horbury et~al.}(2008)\textit{Horbury, Forman, and
  Oughton}}]{horbury08}
Horbury, T., M.~Forman, and S.~Oughton (2008), Anisotropic scaling of
  magnetohydrodynamic turbulence, \textit{Phys. Rev. Lett.}, \textit{101}(17),
  \doi{10.1103/PhysRevLett.101.175005}.

\bibitem[{\textit{Horbury et~al.}(2012)\textit{Horbury, Wicks, and Chen}}]{horbury12}
Horbury, T., R.~Wicks, and C.~Chen (2012), Ansiotropy in Space
Plasma Turbulence: Solar Wind Observations,
\textit{Space Sci. Rev.},
\textit{172}{325}

\bibitem[{\textit{Huang et~al.}(2010)\textit{Huang, Zhou, Sahraoui, Deng, Pang, Yuan, Wei, Wang, and Zhou}}]{huang10}
Huang, S.Y., M.~Zhou, F.~Sahraoui, X.H.~Deng, Y.~Pang, Z.G.~Yuan, Q.~Wei, J.F.~Wang and X.M.~Zhou (2010),
  Wave properties in the magnetic reconnection diffusion region with high $\beta$: Application of the $k$-filtering method to Cluster multispacecraft data, \textit{J. Geophy. Res.}, \textit{115}(A12211),
  \doi{10.1029/2010JA015335}.

\bibitem[{\textit{Hunter}(2007)}]{matplotlib}
Hunter, J.~D. (2007), Matplotlib: A 2d graphics environment, \textit{Computing
  In Science \& Engineering}, \textit{9}(3), 90--95,
  \doi{10.1109/MCSE.2007.55}.

\bibitem[{\textit{{Iroshnikov}}(1964)}]{iroshnikov64}
{Iroshnikov}, P.~S. (1964), {Turbulence of a Conducting Fluid in a Strong
  Magnetic Field}, \textit{Soviet Astronomy}, \textit{7}, 566.

\bibitem[{\textit{Klein et~al.}(2012)}]{klein12}
{Klein}, K., G. {Howes}, J. {TenBarge}, S. {Bale}, C. {Chen}, and C. {Salem} (2012),
  {Using Synthetic Spacecraft Data to Interpret Compressible
  Fluctuations in Solar Wind Turbulence},
  \textit{Astrophys. J.}, \textit{755}, 159.

\bibitem[{\textit{Klein et~al.}(2014)}]{klein14}
{Klein}, K., G. {Howes}, J. {TenBarge}, and J. {Podesta} (2014), {Physical interpretation of the angle-dependent
  magnetic helicity spectrum in the solar wind:
  the nature of turbulent fluctuations near the proton gyroradius scale}, \textit{Astrophys. J.}, \textit{785},
  138, \doi{10.1088/0004-637X/785/2/138}.

\bibitem[{\textit{Klein et~al.}(2019)}]{klein19}
{Klein}, K., O. {Alexandrova}, J. {Bookbinder}, D. {Capriori},
A. {Case}, B. {Chandran}, L. {Chen}, T. {Horbury}, L. {Jian},
J. {Casper}, O. {Le Contel}, B. {Maruca}, W. {Matthaeus},
A. {Retino}, O. {Roberts}, A. {Schekochihin}, R. {Skoug},
C. {Smith}, J. {Steinburg}, H. {Spence}, B. {Vasquez},
J. {TenBarge}, D. {Verscharen}, and P. {Whittlesey}
{[Plasma 2020 Decadal] Multipoint Measurements of the Solar Wind: A Proposed Advance for Studying Magnetized Turbulence},
\textit{arXiv}: 1903.05740.

\bibitem[{\textit{{Mallet} et~al.}}(2015)]{mallet15}
  {Mallet}, A., A. {Schekochihin}, and B. {Chandran} (2015),
  {Refined critical balance in strong Alfv\'enic turbulence},
  \textit{Mon. Not. R. Astron. Soc.}, \textit{449}, L77--L81.

\bibitem[{\textit{{Maron} and {Goldreich}}(2001)}]{maron01}
{Maron}, J., and P.~{Goldreich} (2001), {Simulations of Incompressible
  Magnetohydrodynamic Turbulence}, \textit{Astrophys. J.}, \textit{554},
  1175--1196, \doi{10.1086/321413}.

\bibitem[{\textit{{Matthaeus} et~al.}(1990)\textit{{Matthaeus}, {Goldstein},
  and {Roberts}}}]{matthaeus90}
{Matthaeus}, W.~H., M.~L. {Goldstein}, and D.~A. {Roberts} (1990), {Evidence
  for the presence of quasi-two-dimensional nearly incompressible fluctuations
  in the solar wind}, \textit{J. Geophys. Res.}, \textit{95}, 20,673--20,683,
  \doi{10.1029/JA095iA12p20673}.

\bibitem[{\textit{Matthaeus et~al.}(1996)\textit{Matthaeus, Ghosh, Oughton, and
  Roberts}}]{matthaeus96}
Matthaeus, W.~H., S.~Ghosh, S.~Oughton, and D.~A. Roberts (1996), {Anisotropic
  three-dimensional {MHD} turbulence}, \textit{{J. Geophys. Res.}},
  \textit{101}, \doi{10.1029/95JA03830}.

\bibitem[{\textit{Motschmann et~al.}(1996)\textit{Motschmann, Woodward, Glassmeier, Southwood and {Pin\c con}}}]{motschmann96}
Motschmann, U., T.I.~Woodward, K.H.~Glassmeier, D.J.~Southwood,
  and J.L.~{Pin\c con} (1996),
  Wavelength and Direction Filtering by Magnetic Measurements
  at Satellite Arrays: Generized Minimum Variance Analysis,
  \textit{J. Geophys. Res.}, \textit{101}, 4961--4965.

\bibitem[{\textit{{Narita} and {Glassmeier}}(2009)}]{narita09}
{Narita}, Y., and K.-H. {Glassmeier} (2009),
Spatial aliasing and distortion of energy distribution
in the wave vector domain under multi-spacecraft measurements,
  \textit{Annales Geophysicae},
  \textit{27}, 3031--3042.

\bibitem[{\textit{{Narita} et~al.}(2011)\textit{{Narita}, {Glassmeier}, and
  {Motschmann}}}]{narita11}
{Narita}, Y., K.-H. {Glassmeier}, and U.~{Motschmann} (2011), {High-resolution
  wave number spectrum using multi-point measurements in space - the
  Multi-point Signal Resonator (MSR) technique}, \textit{Annales Geophysicae},
  \textit{29}, 351--360, \doi{10.5194/angeo-29-351-2011}.

\bibitem[{\textit{Narita et~al.}(2014)\textit{Narita, Comisel, and
  Motschmann}}]{narita14}
Narita, Y., H.~Comisel, and U.~Motschmann (2014), Spatial structure of
  ion-scale plasma turbulence, \textit{Frontiers in Physics}, \textit{2}, 13,
  \doi{10.3389/fphy.2014.00013}.

\bibitem[{\textit{{Neubauer} and {Glassmeier}}(1990)}]{neubauer90}
Neubauer, F.~M. and K.-H. Glassmeier (1990),
Use of an Array of Satellites as a Wave Telescope,
  \textit{J. Geophys. Res.}, \textit{95}, 19115--19122.

\bibitem[{\textit{{Paschmann} and {Patrick}}}(1998)]{paschmann98}
Paschmann, G, and D. Patrick (1998),
{Analysis Methods for Multi-Spacecraft Data},
\textit{ISSI Scientific Report Series}.

\bibitem[{\textit{{Paschmann} and {Patrick}}}(2008)]{paschmann08}
Paschmann, G, and D. Patrick (2008),
{Multi-Spacecraft Analysis Method Revisted},
\textit{ISSI Scientific Report Series}.

\bibitem[{\textit{{Pin\c{c}on} and {Lefeuvre}}(1988)}]{pincon88}
{Pin\c{c}on}, J.~L., and F.~{Lefeuvre} (1988),
{Characterization of a homogeneous field turbulence from multipoint measurements},
  \textit{Adv. in Space Res.},
  \textit{8}, 459--462.

\bibitem[{\textit{{Pin\c{c}on} and {Lefeuvre}}(1991)}]{pincon91}
{Pin\c{c}on}, J.~L., and F.~{Lefeuvre} (1991), {Local characterization of
  homogeneous turbulence in a space plasma from simultaneous measurements of
  field components at several points in space}, \textit{{J. Geophys. Res.}},
  \textit{96}, 1789--1802, \doi{10.1029/90JA02183}.

\bibitem[{\textit{{Podesta}}(2009)}]{podesta09}
{Podesta}, J. (2009), {Dependence of Solar-Wind Power Spectra on the Direction
  of the Local Mean Magnetic Field}, \textit{Astrophys. J.}, \textit{698},
  986--999, \doi{10.1088/0004-637X/698/2/986}.

\bibitem[{\textit{{Ramachandran} and {Varoquaux}}(2011)}]{mayavi}
{Ramachandran}, P. and {Varoquaux}, G. (2011), {Mayavi: 3D Visualization of Scientific Data}, \textit{{IEEE Computing in Science \& Engineering}}, \textit{13}, 40--51

\bibitem[{\textit{Retino et~al.}(2019)}]{retino19}
Retino, A. et al.,
{PROSPERO PRObing the Structure of Plasma Energisation RegiOns. Science theme: Exploring the space-time plasma Universe},
\url{https://www.lpp.polytechnique.fr/IMG/pdf/esa-f1-prospero.pdf}.

\bibitem[\textit{{Sahraoui} et~al.}(2003)]{sahraoui03}
{Sahraoui}, F., J.~L. {Pin\c{c}on},
G. {Belmont}, N. {Cornilleau-Wehrlin},
F. {Robert}, L. {Mellul},
J.~M. {Bosqued}, A. {Balogh}, P. {Canu},
and G. {Chanteur} (2003),
  {ULF wave identification in the magnetosheath: The k-filtering technique applied to Cluster II data},
  \textit{J. Geophys. Res.},
  \textit{108}, 1335.

\bibitem[{\textit{{Sahraoui} et~al.}(2010{\natexlab{a}})\textit{{Sahraoui},
  {Goldstein}, {Belmont}, {Canu}, and {Rezeau}}}]{sahraoui10prl}
{Sahraoui}, F., M.~L. {Goldstein}, G.~{Belmont}, P.~{Canu}, and L.~{Rezeau}
  (2010{\natexlab{a}}), {Three Dimensional Anisotropic k Spectra of Turbulence
  at Subproton Scales in the Solar Wind}, \textit{Physical Review Letters},
  \textit{105}(13), 131101, \doi{10.1103/PhysRevLett.105.131101}.

\bibitem[{\textit{{Sahraoui} et~al.}(2010{\natexlab{b}})\textit{{Sahraoui},
  {Belmont}, {Goldstein}, and {Rezeau}}}]{sahraoui10}
{Sahraoui}, F., G.~{Belmont}, M.~L. {Goldstein}, and L.~{Rezeau}
  (2010{\natexlab{b}}), {Limitations of multispacecraft data techniques in
  measuring wave number spectra of space plasma turbulence}, \textit{Journal of
  Geophysical Research (Space Physics)}, \textit{115}, A04206,
  \doi{10.1029/2009JA014724}.

\bibitem[{\textit{{Salem} et~al.}(2012)}]{salem12}
{Salem}, C., G.~{Howes}, D.~{Sundkvist}, S.~{Bale}, C.~{Chaston},
  C.~{Chen}, and F.~{Mozer} (2012),
  {Identification of Kinetic Alfv\'en Wave Turbulence in the Solar Wind},
  \textit{Astrophys. J. Lett.}, \textit{745}, L9.

\bibitem[{\textit{{Samson}}(1983{\natexlab{a}})}]{samson83}
{Samson}, J.~C. (1983{\natexlab{a}}), {Pure states, polarized waves, and
  principal components in the spectra of multiple, geophysical time-series},
  \textit{Geophysical Journal}, \textit{72}, 647--664,
  \doi{10.1111/j.1365-246X.1983.tb02825.x}.

\bibitem[{\textit{{Samson}}(1983{\natexlab{b}})}]{samson83b}
{Samson}, J.~C. (1983{\natexlab{b}}), The reduction of sample-bias in
  polarization estimators for multichannel geophysical data with anisotropic
  noise, \textit{Geophysical Journal International}, \textit{75}(2), 289--308,
  \doi{10.1111/j.1365-246X.1983.tb01927.x}.

\bibitem[{\textit{{Santol\'{\i}k} et~al.}(2003)\textit{{Santol\'{\i}k},
  {Parrot}, and {Lefeuvre}}}]{santolik03}
{Santol\'{\i}k}, O., M.~{Parrot}, and F.~{Lefeuvre} (2003), {Singular value
  decomposition methods for wave propagation analysis}, \textit{Radio Science},
  \textit{38}, 1010, \doi{10.1029/2000RS002523}.

\bibitem[{\textit{{Schmidt}}(1986)}]{schmidt86}
{Schmidt}, R.~O. (1986), {Multiple emitter location and signal parameter
  estimation}, \textit{IEEE Transactions on Antennas and Propagation},
  \textit{34}, 276--280, \doi{10.1109/TAP.1986.1143830}.

\bibitem[{\textit{{Shebalin} et~al.}(1983)\textit{{Shebalin}, {Matthaeus}, and
  {Montgomery}}}]{shebalin83}
{Shebalin}, J.~V., W.~H. {Matthaeus}, and D.~{Montgomery} (1983), {Anisotropy
  in MHD turbulence due to a mean magnetic field}, \textit{Journal of Plasma
  Physics}, \textit{29}, 525--547, \doi{10.1017/S0022377800000933}.

\bibitem[{\textit{{Shi} et~al.}(2015)\textit{{Shi}, {Xiao}, {Li}, {Wang},
  {Wang}, and {LI}}}]{shi15}
{Shi}, M.~J., C.~J. {Xiao}, Q.~S. {Li}, H.~G. {Wang}, X.~G. {Wang}, and H.~{LI}
  (2015), {Observations of Alfv{\'e}n and Slow Waves in the Solar Wind near 1
  AU}, \textit{{Astrophys. J.}}, \textit{815}, 122,
  \doi{10.1088/0004-637X/815/2/122}.

\bibitem[{\textit{{Sonnerup} and {Cahill}}(1967)}]{sonnerup67}
{Sonnerup}, B.~U.~O., and J.~L.~J. {Cahill} (1967), {Magnetopause Structure and
  Attitude from Explorer 12 Observations}, \textit{{J. Geophys. Res.}},
  \textit{72}, 171, \doi{10.1029/JZ072i001p00171}.

\bibitem[{\textit{{Tjulin} et~al.}(2005)\textit{{Tjulin}, {Pin{\c c}On},
  {Sahraoui}, {Andr{\'e}}, and {Cornilleau-Wehrlin}}}]{tjulin05}
{Tjulin}, A., J.-L. {Pin{\c c}On}, F.~{Sahraoui}, M.~{Andr{\'e}}, and
  N.~{Cornilleau-Wehrlin} (2005), {The k-filtering technique applied to wave
  electric and magnetic field measurements from the Cluster spacecrafts},
  \textit{Journal of Geophysical Research (Space Physics)}, \textit{110},
  A11224, \doi{10.1029/2005JA011125}.

\bibitem[\textit{{Tooley} et~al.}(2016)]{tooley2016}
{Tooley}, C.R., R.K.~{Black}, B.P.~{Robertson}, J.M.~{Stone}, S.E.~{Pope}, and G.T.~{Davis},
{The Magnetospheric Multiscale Constellation},
\textit{Space Science Reviews}, \textit{199}, 23--76, \doi{10.1007/s11214-015-0220-5}.

\bibitem[{\textit{{Tu} and Marsch}({1995})}]{tu95}
{Tu}, C., and E.~Marsch ({1995}), {MHD STRUCTURES, WAVES AND TURBULENCE IN THE
  SOLAR-WIND - OBSERVATIONS AND THEORIES}, \textit{{Space. Sci. Rev.}},
  \textit{{73}}({1-2}), 1--210, \doi{10.1007/BF00748891}.

\bibitem[{\textit{{Turner} et~al.}(2012)\textit{{Turner}, {Gogoberidze}, and
  {Chapman}}}]{turner12}
{Turner}, A.~J., G.~{Gogoberidze}, and S.~C. {Chapman} (2012), {Nonaxisymmetric
  Anisotropy of Solar Wind Turbulence as a Direct Test for Models of
  Magnetohydrodynamic Turbulence}, \textit{Physical Review Letters},
  \textit{108}(8), 085001, \doi{10.1103/PhysRevLett.108.085001}.

\bibitem[{\textit{Verscharen et~al.}(2017)}]{verscharen17}
  {Verscharen, D., C.~{Chen}, and R.~{Wicks}} (2017),
  {On Kinetic Slow Modes, Fluid Slow Modes, and Pressure-balanced Structures in the Solar Wind},
  \textit{Astrophys. J. }, \textit{840}, 106.

\bibitem[{\textit{Verscharen et~al.}(2019)}]{verscharen19}
  {Verscharen, D., R.~{Wicks}, O.~{Alexandrova}, R.~{Bruno},  et~al.} (2019),
  {A Case For Electron Astrophyscis},
  \textit{arXiv}, 1908.02206

\bibitem[{\textit{Vestuto et~al.}({2003})\textit{Vestuto, Ostriker, and
  Stone}}]{vestuto03}
Vestuto, J., E.~Ostriker, and J.~Stone ({2003}), {Spectral properties of
  compressible magnetohydrodynamic turbulence from numerical simulations},
  \textit{{Astrophys. J.}}, \textit{{590}}({2, 1}), 858--873,
  \doi{10.1086/375021}.

\bibitem[{\textit{{Wang} et~al.}(2012)\textit{{Wang}, {He}, {Tu}, {Marsch},
  {Zhang}, and {Chao}}}]{wang12}
{W{}ang}, X., J.~{He}, C.~{Tu}, E.~{Marsch}, L.~{Zhang}, and J.-K. {Chao}
  (2012), {Large-amplitude Alfv{\'e}n Wave in Interplanetary Space: The Wind
  Spacecraft Observations}, \textit{{Astrophys. J.}}, \textit{746}, 147,
  \doi{10.1088/0004-637X/746/2/147}.

\bibitem[{\textit{Wang et~al.}(2016)\textit{Wang, Cao, Fu, Meng, and Dunlop}}]{wang16}
Wang, T.Y., J.B.~Cao, H.S.~Fu, X.J.~Meng, and M.~Dunlop (2016),
  Compressible turbulence with slow-mode waves observed in the bursty bulk flow of plasma sheet, \textit{Geophys. Res. Lett.}, \textit{43}(1854),
  \doi{10.1002/2016GL068147}.

\bibitem[{\textit{Wicks et~al.}(2019)}]{debyemission}
Wicks, R.T. et al., Debye Mission Proposal,
  \url{www.ucl.ac.uk/mssl/research-projects/2019/may/debye}.

\bibitem[{\textit{{Yan} et~al.}(2016)\textit{{Yan}, {He}, {Zhang}, {Tu},
  {Marsch}, {Chen}, {Wang}, {Wang}, and {Wicks}}}]{yan16}
{Yan}, L., J.~{He}, L.~{Zhang}, C.~{Tu}, E.~{Marsch}, C.~H.~K. {Chen},
  X.~{Wang}, L.~{Wang}, and R.~T. {Wicks} (2016), {Spectral Anisotropy of
  Els{\"a}sser Variables in Two-dimensional Wave-vector Space as Observed in
  the Fast Solar Wind Turbulence}, \textit{Astropys. J. Lett.}, \textit{816},
  L24, \doi{10.3847/2041-8205/816/2/L24}.

\bibitem[{\textit{Yao et~al.}(2011)\textit{Yao, He, Marsch, Tu, Pedersen,
  R{\`e}me, and Trotignon}}]{yao11}
Yao, S., J.~He, E.~Marsch, C.~Tu, A.~Pedersen, H.~R{\`e}me, and J.~G. Trotignon
  (2011), {{MULTI-SCALE} {ANTI-CORRELATION} {BETWEEN} {ELECTRON} {DENSITY}
  {AND} {MAGNETIC} {FIELD} {STRENGTH} {IN} {THE} {SOLAR} {WIND}},
  \textit{{Astrophys. J.}}, \textit{728}, \doi{10.1088/0004-637X/728/2/146}.

\bibitem[{\textit{Zhang et~al.}(2020)}]{dataset_S1_S2}
Zhang, L., J.~He, Y.~Narita, and X.~S. Feng (2020), 
{Supplymented Data for ``On the Conditions of Aliasing Effects in Constellations with More than Four Spacecraft''}. 
Zenodo. \url{http://doi.org/10.5281/zenodo.4146803}. 
Accessed 2020-11-03. 

\end{thebibliography}

\end{document}